\documentclass{article}

\usepackage{graphicx}
\usepackage{epstopdf}
\usepackage{mdframed}
\usepackage{tcolorbox}
\usepackage{lipsum}
\usepackage[backend=biber, style=numeric-comp, sorting=none, natbib, maxbibnames=99, maxcitenames=2, uniquename=false, uniquelist=false]{biblatex}
\DeclareNameAlias{sortname}{last-first}
\DeclareNameAlias{default}{last-first}
\usepackage{amsmath}
\usepackage[font=small,labelfont=bf,labelsep=space]{caption} 
\usepackage{subcaption}
\usepackage{float}
\usepackage{mathtools}
\usepackage{amssymb}
\usepackage{bm}
\usepackage[margin=1.35in]{geometry}
\usepackage{rotating}
\usepackage[para]{threeparttable}

\usepackage{booktabs} 
\usepackage{multirow}
\usepackage{gensymb}
\usepackage{enumitem}
\setlist[itemize]{noitemsep, topsep=0pt}
\usepackage[utf8]{inputenc}
\usepackage[T1]{fontenc}
\usepackage{xcolor}

\definecolor{col1}{rgb}{0.      , 0.135112, 0.304751}
\definecolor{col2}{rgb}{0.      , 0.160495, 0.364534}
\definecolor{col3}{rgb}{0.      , 0.18455 , 0.428802}
\definecolor{col4}{rgb}{0.068968, 0.209372, 0.438863}
\definecolor{col5}{rgb}{0.13683 , 0.234216, 0.432148}
\definecolor{col6}{rgb}{0.190303, 0.261644, 0.426329}
\definecolor{col7}{rgb}{0.230871, 0.286134, 0.423498}
\definecolor{col8}{rgb}{0.271639, 0.313253, 0.422837}
\definecolor{col9}{rgb}{0.305886, 0.337681, 0.424512}
\definecolor{col10}{rgb}{0.342246, 0.364939, 0.428559}
\definecolor{col11}{rgb}{0.373884, 0.389646, 0.434209}
\definecolor{col12}{rgb}{0.408226, 0.417357, 0.44257}
\definecolor{col13}{rgb}{0.438504, 0.44258 , 0.452341}
\definecolor{col14}{rgb}{0.471501, 0.47096 , 0.466357}
\definecolor{col15}{rgb}{0.503185, 0.496851, 0.472305}
\definecolor{col16}{rgb}{0.540307, 0.526005, 0.472163}
\definecolor{col17}{rgb}{0.574417, 0.552682, 0.469172}
\definecolor{col18}{rgb}{0.612977, 0.582861, 0.46295}
\definecolor{col19}{rgb}{0.648222, 0.610553, 0.454801}
\definecolor{col20}{rgb}{0.687957, 0.641966, 0.442886}
\definecolor{col21}{rgb}{0.724274, 0.670859, 0.429194}
\definecolor{col22}{rgb}{0.765223, 0.703705, 0.410587}
\definecolor{col23}{rgb}{0.802667, 0.733978, 0.390153}
\definecolor{col24}{rgb}{0.844957, 0.76845 , 0.362741}
\definecolor{col25}{rgb}{0.88372 , 0.800258, 0.332599}
\definecolor{col26}{rgb}{0.927724, 0.836486, 0.290611}
\definecolor{col27}{rgb}{0.968469, 0.869819, 0.24131}
\definecolor{col28}{rgb}{0.995737, 0.909344, 0.217772}

\definecolor{civ_col1}{rgb}{0.000000, 0.135112, 0.304751}
\definecolor{civ_col2}{rgb}{0.165113, 0.247965, 0.428908}
\definecolor{civ_col3}{rgb}{0.342246, 0.364939, 0.428559}
\definecolor{civ_col4}{rgb}{0.488697, 0.485318, 0.471008}
\definecolor{civ_col5}{rgb}{0.648222, 0.610553, 0.454801}
\definecolor{civ_col6}{rgb}{0.823729, 0.751101, 0.377043}
\definecolor{civ_col7}{rgb}{0.995737, 0.909344, 0.217772}
\usepackage{pifont}
\newcommand{\cmark}{\ding{51}}
\newcommand{\xmark}{\ding{55}}%
\usepackage[bottom]{footmisc}
\usepackage[normalem]{ulem}
\usepackage{dirtytalk}
\usepackage{setspace}
\usepackage{upgreek}
\usepackage[makeroom]{cancel}
\usepackage{nomencl}
\makenomenclature
\usepackage{tikz}
\usepackage{pgfplotstable}
\pgfplotsset{compat=1.17}
\newcommand{\irow}[1]{%
  \begin{smallmatrix}(#1)\end{smallmatrix}%
}
\allowdisplaybreaks
\usepackage{lettrine}
\newcommand*\dropcap[2]{
    \lettrine[lines=2,findent=0.2em,nindent=0pt]{#1}{#2}%
}
\usepackage{hyperref}
\hypersetup{bookmarks=true,breaklinks=true}
\usepackage[xindy={glsnumbers=false}, nonumberlist, nogroupskip]{glossaries}
\input{glossary.cls}
\makeglossaries
\newenvironment{myabs}[1]
    {\begin{center}
        \begin{tcolorbox}[width=0.75\linewidth,arc=0pt,colback=gray!20,frame empty]
            \begin{center}
                \textbf{ABSTRACT}
            \end{center}
            \smallbreak
            }
            {
        \end{tcolorbox}
    \end{center}
    }

\title{ON FREQUENCY-DEPENDENT ROCK EXPERIMENTS: A COMPARATIVE REVIEW}

\author{\href{mailto:stian.rorheim@ntnu.no}{Stian R{\o}rheim}\thanks{\acrfull{ntnu}, Trondheim, Norway.}}

\date{August $7$, $2022$}

\begin{document}

\maketitle


\begin{myabs}

\dropcap{\textbf{R}}{ock} properties are environment- and condition-dependent which render field-laboratory comparisons ambiguous for a number of known and unknown reasons that constitute the upscaling problem. Unknowns are first transformed into knowns in a controlled environment (laboratory) and second in a volatile environment (field). Causality-bound dispersion and attenuation are respectively defined as rock properties that are frequency- and distance-dependent: dispersion implies non-zero attenuation and vice versa. \acrfull{fo}, \acrfull{rb}, and \acrfull{pt} are the customary techniques to measure rock properties at Hz, kHz, and MHz frequencies. Notably \acrshort{fo} has emerged as the current champion in bridging the field-laboratory void in recent years. Not only is \acrshort{fo} probing seismic (Hz) frequencies but with $\sim10^{-6}$ strain amplitudes it is also similar to field seismic. \acrshort{rb} and \acrshort{pt} are concisely however \acrshort{fo} is verbosely elaborated by chronologically compiling most (if not all) \acrshort{fo} studies on sedimentary rocks and comparing all available \acrshort{fo} measurements on reference materials such as lucite, aluminium, and \acrshort{peek}. First of its kind, this inter-laboratory comparison may serve as a reference for others who seek to verify their own results. Differences between \acrshort{fo} are discussed with alternative strain and stress sensors being the focal points. Other techniques such as \acrfull{rus}, \acrfull{lus}, and \acrfull{dars} that are similar to \acrshort{fo}, \acrshort{rb}, and \acrshort{pt} are also described. Only time will tell what the future holds for \acrshort{fo} but plausible improvements for the future are ultimately given which may elevate it even further. Experimental combined with numerical novelty will extend the probeable frequencies beyond their current limits.
    
\end{myabs}

\vfill



\clearpage


\section{INTRODUCTION}

Laboratory and field measurements are not easily compared for a number of known and unknown reasons that constitute the upscaling problem. These measurements are based on wave reflection and refraction (field) or transmission and to some extent echo (laboratory). Due to the dispersive (frequency-dependent) nature of fluid-saturated rocks, traditional laboratory (MHz) and field (Hz and kHz) measurements are incompatible. Dispersion is both a laboratory and field phenomenon that occurs whenever properties measured at different frequencies are compared. For example, \citet{white1983} inadvertently described dispersion from seismic to sonic field data while studying anisotropy. Caution must thus be exercised when comparing data or basing models (measured or valid at certain frequencies) on dispersive parameters (measured at other frequencies). Attenuation effects are included in many algorithms for waveform modeling, imaging, and full-waveform inversion among other field-relevant applications. Experimental studies in controlled conditions are paramount to universally understand this fluid-related phenomenon and its mechanisms.

\acrfull{fo}, \acrfull{rb}, and \acrfull{pt} are three common techniques to measure rock properties at seismic (Hz), sonic (kHz), and ultrasonic (MHz) frequencies, respectively. Particularly \acrshort{fo} and to some extent also \acrshort{rb} are increasingly recognized as means to cover seismic and sonic frequencies (from Hz to kHz) in the laboratory. These techniques share the common denominator that all were first used to study metals by physicists but were later adopted by rock physicists to also study rocks. Progress is inevitable as the number of laboratory studies at similar frequencies as in the field are rapidly increasing despite measuring dissimilar parameters. In fact, also the measured parameters are environment-dependent. For example, \acrshort{fo} in its basic (longitudinal) form measures \citeauthor{young1807}'s modulus and \citeauthor{poisson1827}'s ratio instead of P- and S-wave velocities at seismic frequencies, while its auxiliary (uniaxial) form  measures uniaxial modulus equivalent to P-wave velocity. Transformation between moduli and velocities inevitably introduces errors despite being trivial for isotropic and non-trivial for anisotropic rocks. \acrshort{fo} novelty is driven by extending the frequencies at which it operates, what properties it measures and how in terms of sensor types, improving its boundary conditions (dead volume), and recently also its concurrent imaging via \acrshort{ct}.

This study is in its entirety based on \citet{rorheim2022phd} but with minor modifications for formatting purposes. \citet{rorheim2017} was its first iteration and \citet{rorheim2019} its second but it has since been expanded into greater detail while being continuously updated to keep up with time as novel experiments are becoming customary. In fact, unknown to \citet{rorheim2017} at the time, \citet{subramaniyan2014} is an excellent review of apparatuses with the ability to measure seismic attenuation in reservoir rocks. Similarities between this study and \citet{subramaniyan2014} are evident as both are inter-laboratory comparisons however the focus of \citet{subramaniyan2014} was on the standardization need of \acrshort{fo} apparatuses whereas the present is on comparing \acrshort{fo} measured moduli and attenuations on lucite, aluminium, and \acrshort{peek} reference specimens. \citet{subramaniyan2014} did not cover all apparatuses and only considered reservoir rocks such as sandstones and carbonates unlike unconventional ones like shale. Validating attenuation measurements by literature comparison was the basis of \citet{rorheim2019}. Like \citet{subramaniyan2014}, the three most common techniques (\acrshort{fo}, \acrshort{rb}, and \acrshort{pt}) are the focal points (\acrshort{fo} more so than \acrshort{rb} and \acrshort{pt}), but unlike \citet{subramaniyan2014}, other novel techniques such as \acrfull{rus}, \acrfull{dars}, and \acrfull{lus} are also elaborated. The primary three techniques (\acrshort{fo}, \acrshort{rb}, and \acrshort{pt}) are also theoretically defined whereas the secondary ones (\acrshort{rus}, \acrshort{dars}, and \acrshort{lus}) are not. Recent numerical advances are also included as the value and significance of \acrfull{drp} surely will improve in the future when elevated from its current infancy. 


\section{THEORY}

\acrshort{rb}, \acrshort{pt}, and \acrshort{fo} are related due to all three exploiting mechanical disturbances in a material to deduce its properties. Here the theory of \acrshort{fo} is elaborated while that of \acrshort{rb} and \acrshort{pt} are mostly unelaborated due to \acrshort{fo} being the primary focus from the outset and throughout. Relevant models commonly used to simulate the dispersive (or frequency-dependent) behaviour of rocks across frequencies are also briefly explained. 

\subsection{MECHANICAL EQUATIONS}

\acrshort{rb} exploits harmonic waves in the axial direction with radial $u_\mathrm{r}$, circumferential $u_\uptheta$, and axial $u_\mathrm{z}$ components of displacement. $z$-dependent but $\theta$-independent motions are separated into torsional waves with $u_\uptheta$ and longitudinal waves with $u_\mathrm{r}$ and $u_\mathrm{z}$. Flexural waves are $z$- and $\theta$-dependent motions. Phase velocities $V$ at resonating frequencies $f$ relate to bar length $L$ as

\nomenclature{$u_\mathrm{z}$}{Axial displacement.}
\nomenclature{$u_\mathrm{r}$}{Radial displacement.}
\nomenclature{$u_\uptheta$}{Circumferential displacement.}

\begin{equation}
    V=\underbrace{\frac{2L}{n}}_{\lambda}f,
\end{equation}

\nomenclature{$V$}{Velocity.}
\nomenclature{$L$}{Specimen length.}
\nomenclature{$\lambda$}{Wavelength.}
\nomenclature{$f$}{Resonant frequency.}
\nomenclature{$n$}{Positive integer in which $n\in\{1,2,3,\dots\}$ representing nodes or harmonics.}

\noindent where $n\in\{1,2,3,\dots\}$ denotes different modes or harmonics and $\lambda$ is the wavelength. $n=1$ is the fundamental frequency. \citeauthor{young1807}'s modulus $E$ from longitudinal and flexural modes is $E=\rho V_\mathrm{E}^2$, whereas shear modulus $G$ from the torsional mode is $G=\rho V_\mathrm{G}^2$, with $V_\mathrm{E}$, $V_\mathrm{G}$, and $\rho$ being longitudinal or flexural velocity, torsional velocity, and density \citep{bourbie1985}. \acrshort{pt} exploits the time of flight principle to determine P- and S-wave velocities

\nomenclature{$E$}{\citeauthor{young1807}'s modulus.}
\nomenclature{$G$}{Shear modulus.}
\nomenclature{$\rho$}{Density.}
\nomenclature{$V_\mathrm{E}$}{Longitudinal or flexural velocity.}
\nomenclature{$V_\mathrm{G}$}{Torsional velocity.}

\begin{equation}
    V=\frac{L}{\Delta t},
\end{equation}

\begin{figure}[H]
    \centering
    \begin{subfigure}[b]{0.49\textwidth}
        \includegraphics[width=1\linewidth]{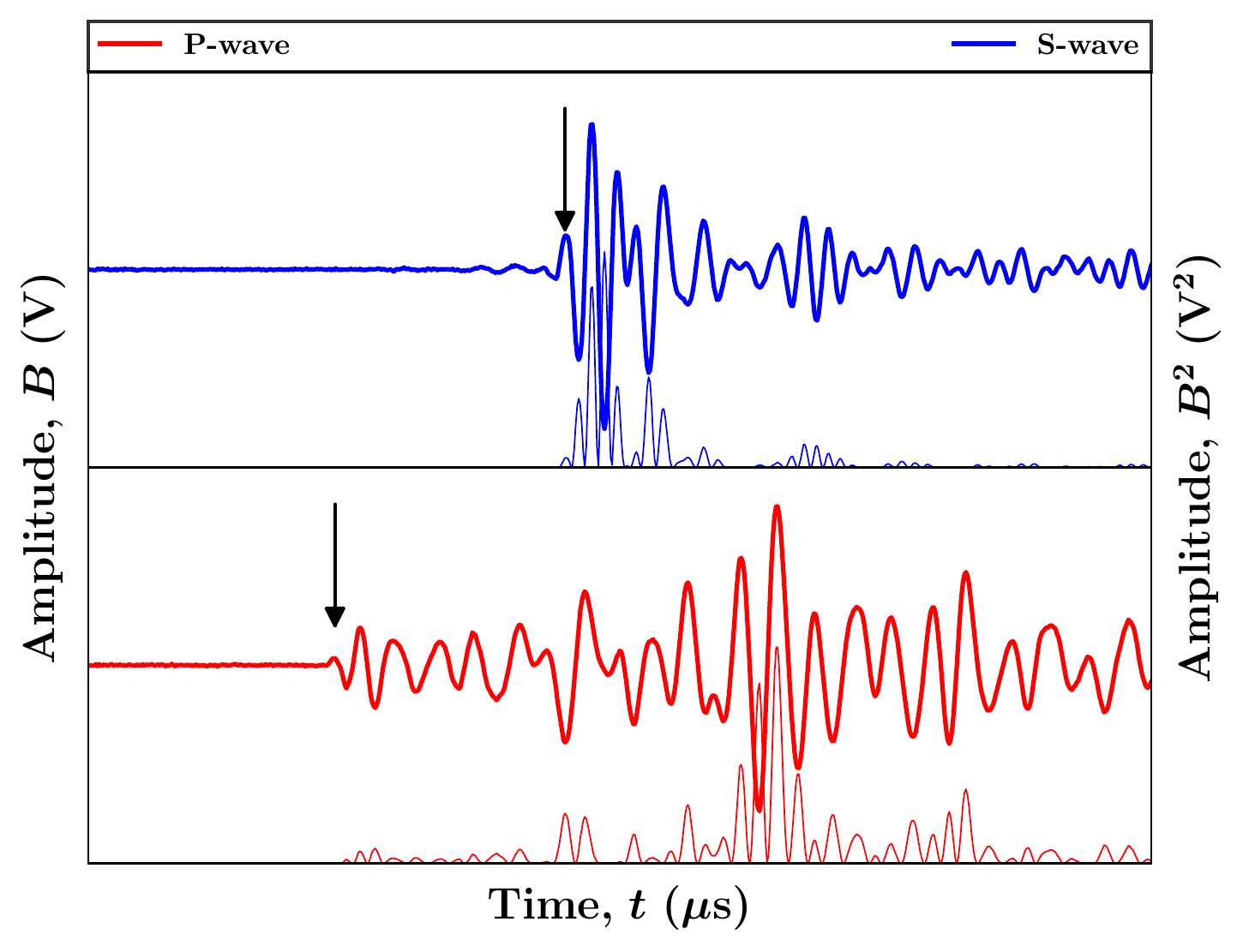}
        \caption{}
        \label{fig:pt} 
    \end{subfigure}
    
    \begin{subfigure}[b]{0.49\textwidth}
        \includegraphics[width=1\linewidth]{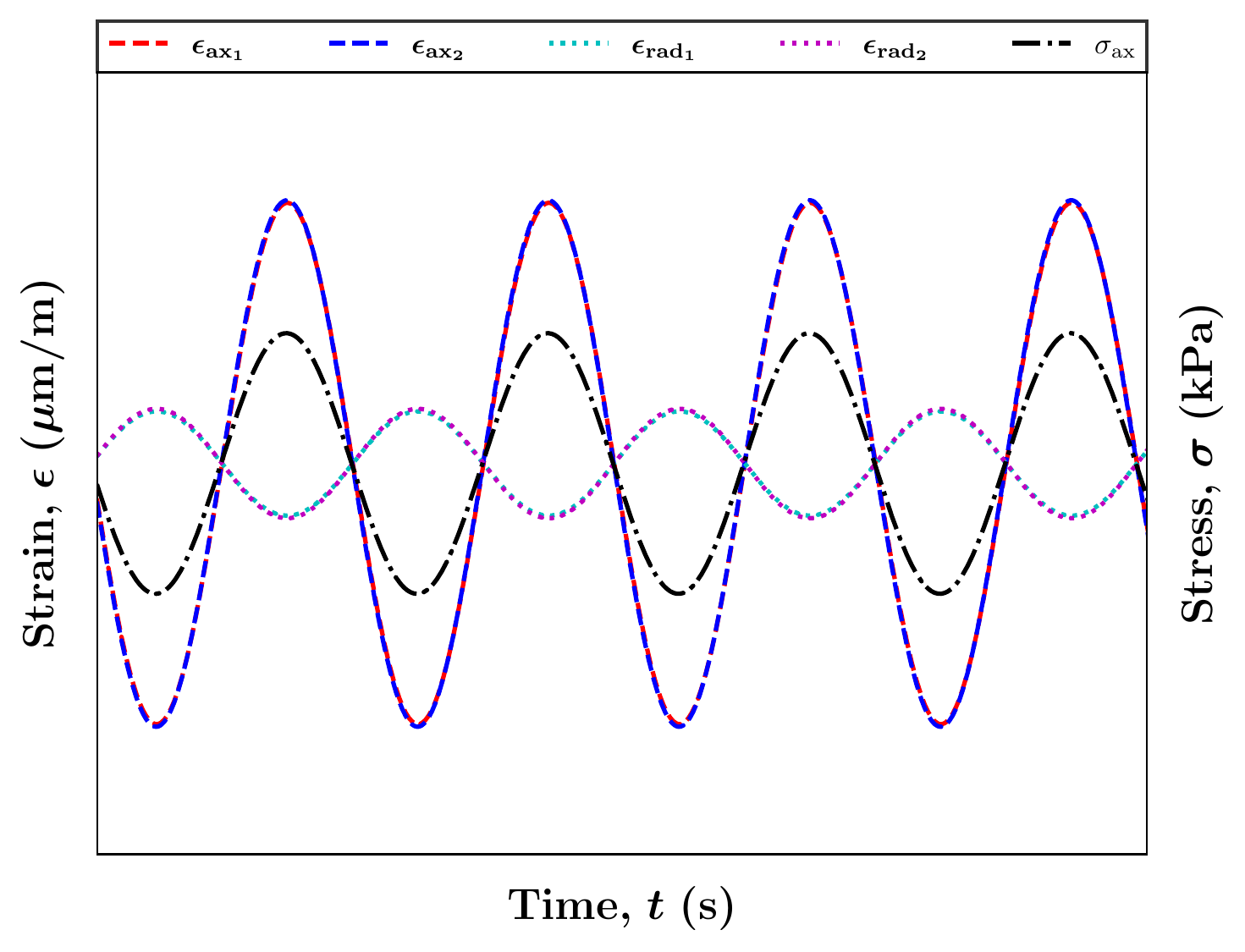}
        \caption{}
        \label{fig:fo-e}
    \end{subfigure}
    \begin{subfigure}[b]{0.49\textwidth}
        \includegraphics[width=1\linewidth]{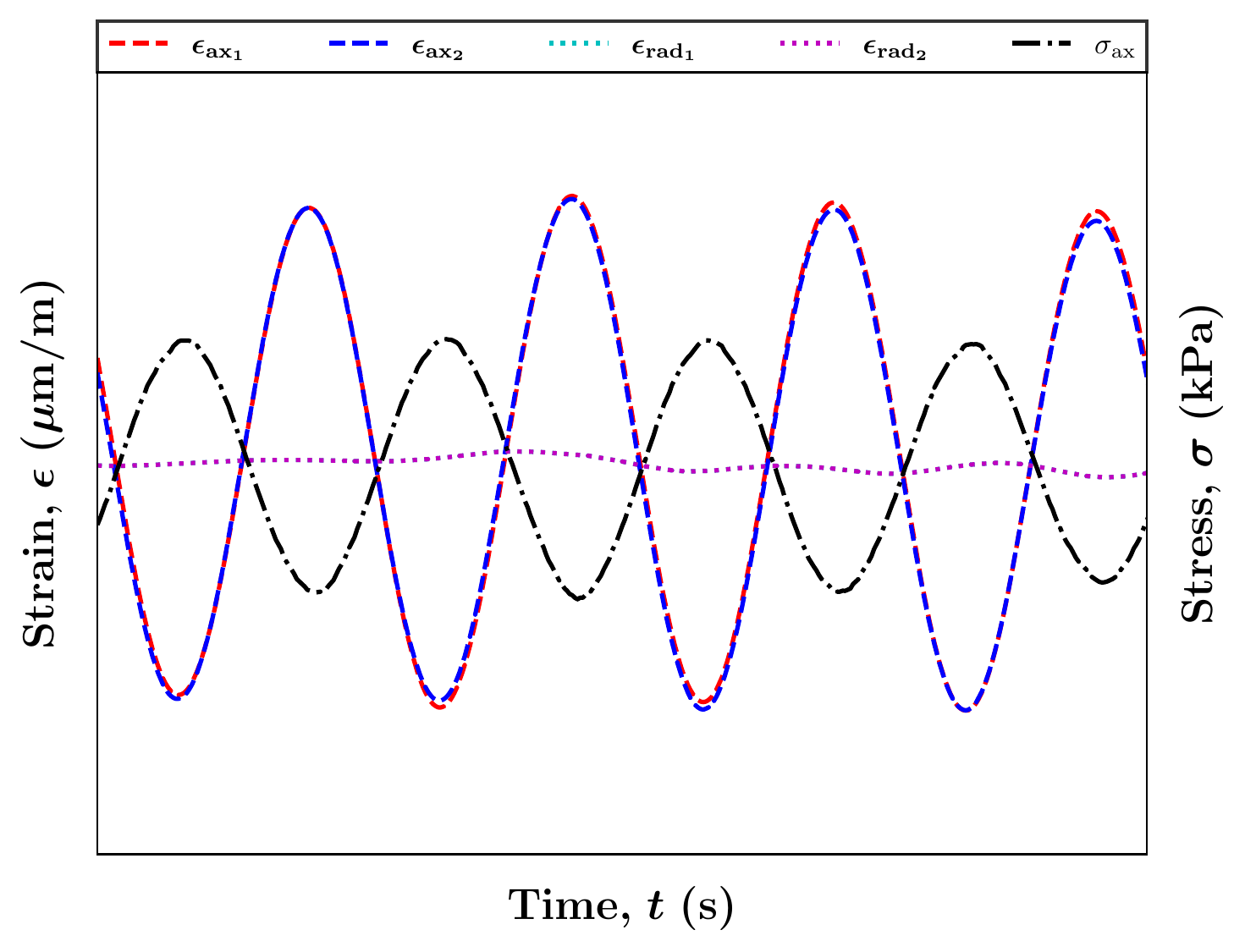}
        \caption{}
        \label{fig:fo-p}
    \end{subfigure}
\caption{Amplitudes $B$, $\epsilon$, and $\sigma$ versus time $t$ exemplified for three different techniques: (\subref{fig:pt}) \acrshort{pt} and two \acrshort{fo} versions: uniaxial-stress (\subref{fig:fo-e}) and uniaxial-strain (\subref{fig:fo-p}). Arrows in (\subref{fig:pt}) indicate first maxima.}
\label{fig:fo-ep-pt}
\end{figure}

\nomenclature{$\Delta t$}{Travel time.}

\noindent where $V$ is the velocity of either body wave type, $L$ is the specimen length, and $\Delta t$ is the travel time (Figure~\ref{fig:pt}). \acrshort{fo} measures different moduli and ratios depending on excitation modes: (i) longitudinal, (ii) torsional, (iii) flexural, (iv) volumetric, and (i) uniaxial. Higher harmonics are also explorable for \acrshort{fo} like for \acrshort{rb}. Elastic moduli are generally defined as stress over strain $M\equiv\sigma/\epsilon$. (i) yields \citeauthor{young1807}'s modulus and \citeauthor{poisson1827}'s ratio $\nu$ respectively as

\nomenclature{$\sigma$}{Stress.}
\nomenclature{$\epsilon$}{Strain.}

\begin{align}
    E   &=\frac{\sigma_\mathrm{ax}}{\epsilon_\mathrm{ax}}, \label{eq:young}\\
    \nu &=-\frac{\epsilon_\mathrm{rad}}{\epsilon_\mathrm{ax}},
\end{align}

\nomenclature{$\sigma_\mathrm{ax}$}{Axial stress.}
\nomenclature{$\epsilon_\mathrm{ax}$}{Axial strain.}
\nomenclature{$\epsilon_\mathrm{rad}$}{Radial strain.}
\nomenclature{$\nu$}{\citeauthor{poisson1827}'s ratio.}

\noindent with $\sigma$ and $\epsilon$ being stress and strain either in \textbf{ax}(ial) or \textbf{rad}(ial) direction (Figure~\ref{fig:fo-e}). (ii) provides shear modulus $G$ as the shear stress-strain ratio 

\begin{equation}
    G = \frac{\sigma_{s}}{\epsilon_{s}},
\end{equation}

\nomenclature{$\sigma_\mathrm{s}$}{Shear stress.}
\nomenclature{$\epsilon_\mathrm{s}$}{Shear strain.}

\noindent in which $\sigma_\mathrm{s}$ and $\epsilon_\mathrm{s}$ are shear stress and shear strain, respectively. \citet{kelvin1856} regarded $K$ and $G$ as principal elasticities with special significance. (iii) adds flexural modulus $E_\mathrm{F}$ but remains undefined due to its condition-dependent definitions. Flexural and \citeauthor{young1807}'s moduli are theoretically equivalent $E=E_\mathrm{F}$ but practically inequivalent $E\neq E_\mathrm{F}$. (iv) grants bulk modulus $K$ as the three-dimensional extension of \citeauthor{young1807}'s modulus $E$. It is defined as an object's proclivity to deform in all directions under hydrostatic stress or pressure regimes

\begin{equation}
    K = \frac{\sigma_\mathrm{p}}{\epsilon_\mathrm{vol}},
\label{eq:bulk}
\end{equation}

\nomenclature{$K$}{Bulk modulus.}
\nomenclature{$\sigma_\mathrm{p}$}{Hydrostatic stress (or pressure).}
\nomenclature{$\epsilon_\mathrm{vol}$}{Volumetric strain.}

\noindent where $\sigma_\mathrm{p}$ (or $P_\mathrm{p}$) is hydrostatic stress (or pressure) and $\epsilon_\mathrm{vol}$ is volumetric strain. (v) gives uniaxial (or P-wave) modulus $H$ which is akin to \citeauthor{young1807}'s modulus $E$ but for $\epsilon_\mathrm{rad}=0$ enforcing uniaxial-strain instead of uniaxial-stress conditions (Figure~\ref{fig:fo-p})

\begin{equation}
    H = \frac{\sigma_\mathrm{ax}}{\epsilon_\mathrm{ax}}\bigg|_{\epsilon_\mathrm{rad}=0}.
    \label{eq:direct_p}
\end{equation}

\nomenclature{$H$}{Uniaxial (or P-wave) modulus.}

\citeauthor{lame1833}'s $\lambda_\mathrm{L}$ and $\mu_\mathrm{L}$ \citep{lame1833} parameterize $K=\lambda_\mathrm{L}+2/3\mu_\mathrm{L}$ where $\mu_\mathrm{L}=G$ and $\lambda_\mathrm{L}$ is physically equivocal. P- and S-wave velocities $V_\mathrm{P}$ and $V_\mathrm{S}$ are then 

\nomenclature{$\lambda_\mathrm{L}$}{\citeauthor{lame1833}'s first parameter.}
\nomenclature{$\mu_\mathrm{L}$}{\citeauthor{lame1833}'s second parameter.}

\begin{align*}
    V_\mathrm{P} &= \sqrt{\frac{K+\frac{4}{3}G}{\rho}}=\sqrt{\frac{\lambda_\mathrm{L}+2\mu_\mathrm{L}}{\rho}},\\
    V_\mathrm{S} &= \sqrt{\frac{G}{\rho}}=\sqrt{\frac{\mu_\mathrm{L}}{\rho}},
\end{align*}

\noindent whereas $V_\mathrm{P}/V_\mathrm{S}=\sqrt{K/G+4/3}=\sqrt{\lambda_\mathrm{L}/\mu_\mathrm{L}+2}$ is density-independent.

\nomenclature{$V_\mathrm{P}/V_\mathrm{S}$}{Ratio between compressional (P) and shear (S) wave velocities.}

\nomenclature{$C_{33}$}{One of five stiffnesses assuming \acrshort{ti} symmetry.}

\subsubsection{ISOTROPY VERSUS ANISOTROPY}

\begin{figure}[H]
    \centering
    \includegraphics[width=0.26\textwidth]{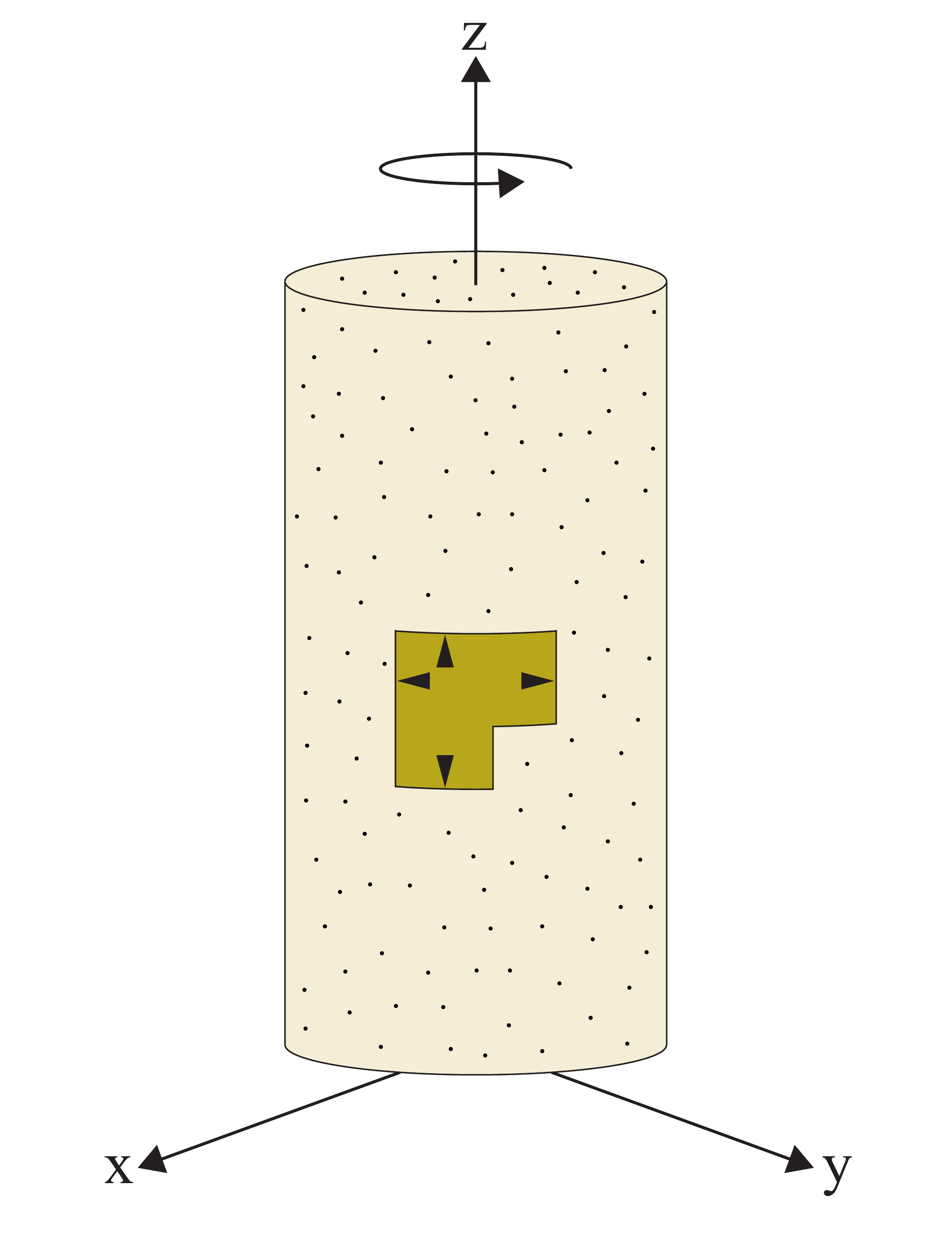}
    \caption{Geometry of an isotropic specimen. Triangles indicate measurement directions of biaxial strain gauges.}
    \label{fig:geo_iso}
\end{figure}

\begin{figure}[H]
    \centering
    \begin{subfigure}[b]{0.26\textwidth}
        \includegraphics[width=1\linewidth]{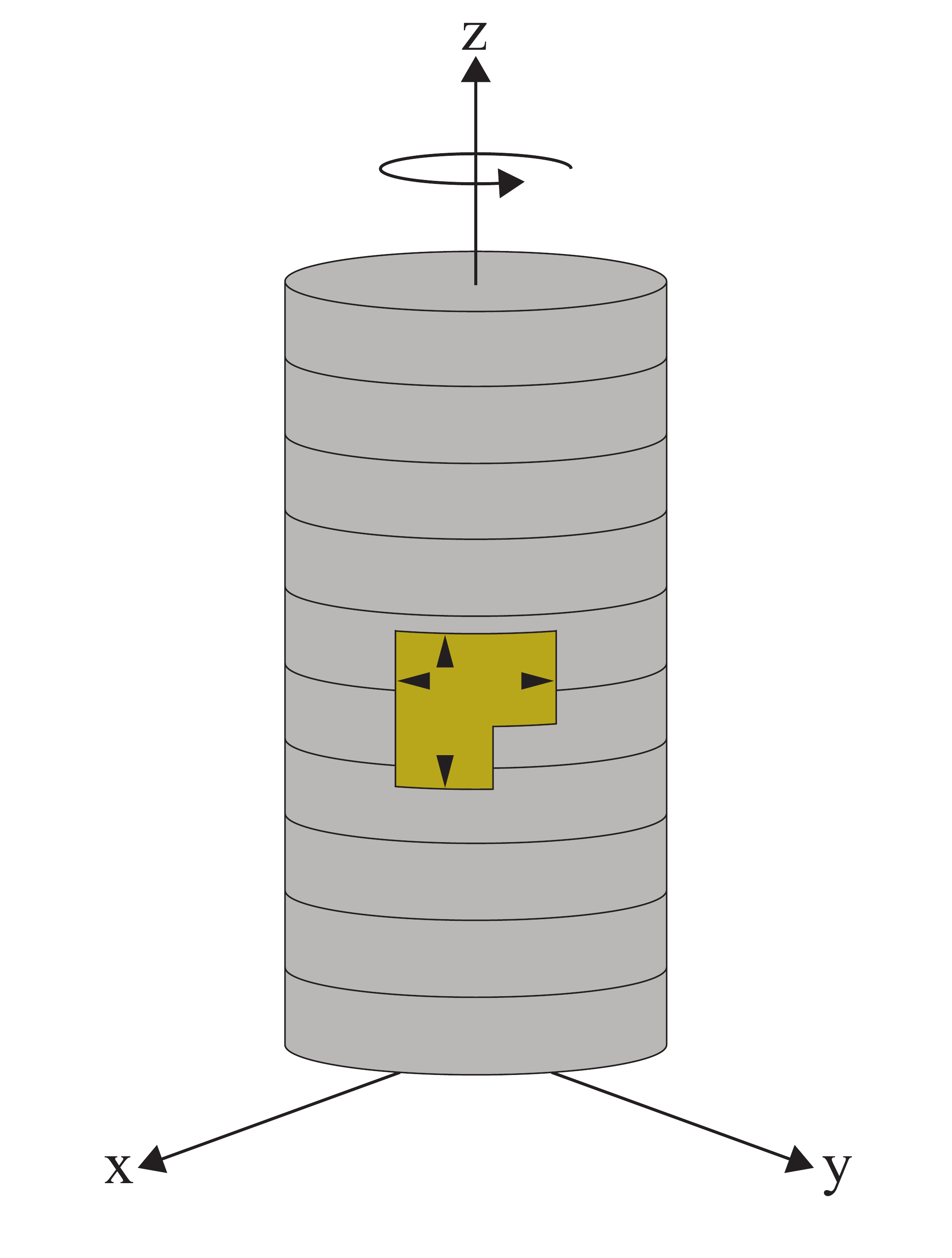}
        \caption{$0\degree$ specimen.}
        \label{fig:geo_aniso_0} 
    \end{subfigure}
    \begin{subfigure}[b]{0.26\textwidth}
        \includegraphics[width=1\linewidth]{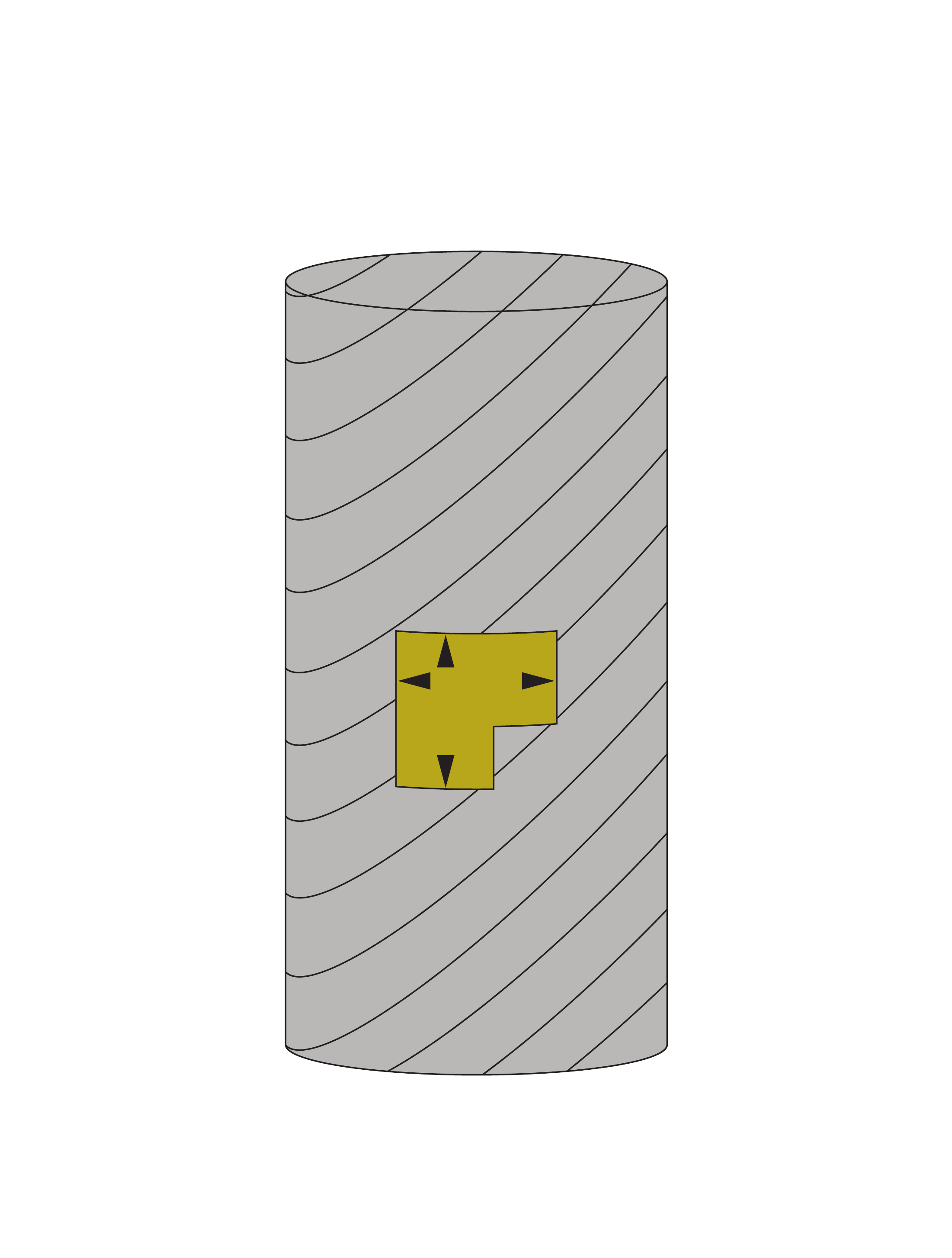}
        \caption{$45\degree$ specimen.}
        \label{fig:geo_aniso_45}
    \end{subfigure}
    \begin{subfigure}[b]{0.26\textwidth}
        \includegraphics[width=1\linewidth]{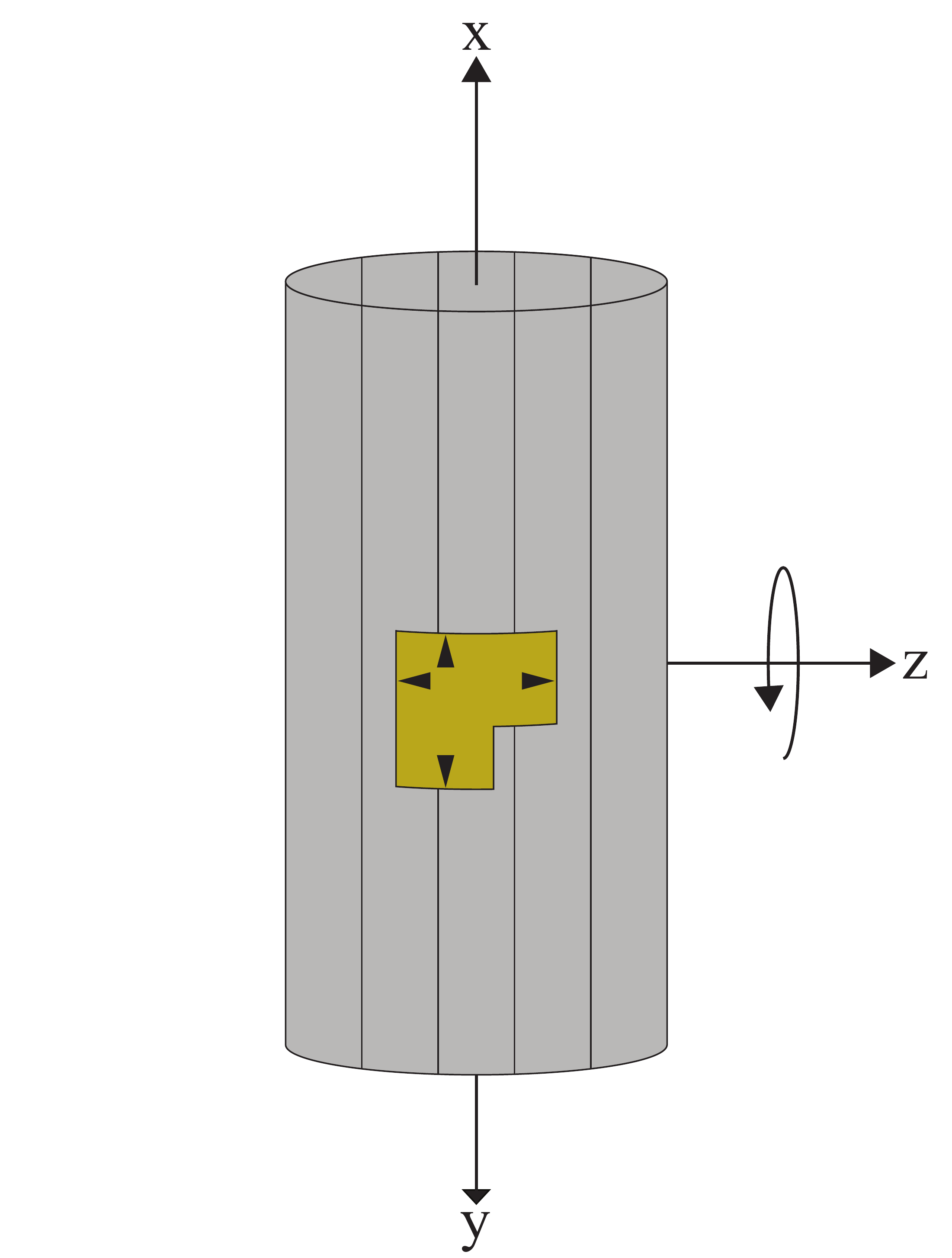}
        \caption{$90\degree$ specimen.}
        \label{fig:geo_aniso_90}
    \end{subfigure}
\caption{Geometries of (\subref{fig:geo_aniso_0}) $0$, (\subref{fig:geo_aniso_45}) $45$, and (\subref{fig:geo_aniso_90}) $90\degree$ specimens assuming \acrshort{ti} symmetry. Triangles indicate measurement directions of biaxial strain gauges.}
\label{fig:geo_aniso}
\end{figure}

Anisotropic or isotropic is the medium whose elastic properties change or unchange with direction. \citeauthor{hooke1678}an theory \citep{hooke1675,hooke1678} first formulated as \say{ut tensio, sic vis}\footnote{\say{As the extension, so the force} is \citeauthor{hooke1678}'s \citeyear{hooke1678} solution \citep{hooke1678} to his \citeyear{hooke1675} anagram \say{ceiiinossssttuu} \citep{hooke1675}.} relates stress $\sigma_{ij}$ and strain $\epsilon_{kl}$ as $\sigma_{ij}=\sum_{k=1}^{3}\sum_{l=1}^{3}C_{ijkl}\epsilon_{kl}$ where $C_{ijkl}$ is stiffness and $i,j=1,2,3$ using \citeauthor{einstein1916} notation \citep{einstein1916}. Transforming between $E$, $\nu$, $G$, $K$, and $H$ is isotropically trivial (Table~\ref{tab:elastic_relations}) yet anisotropically non-trivial as the number of specimens and stiffnesses required for full description increases from one (Figure~\ref{fig:geo_iso}) to three (Figure~\ref{fig:geo_aniso}) and from two (Equation~\ref{eq:iso_cij_matrix}) to five (Equation~\ref{eq:aniso_cij_matrix}) using \citeauthor{voigt1910} notation \citep{voigt1910} for \acrfull{ti}, respectively. Stress $\bm{\overrightarrow{\sigma}}=\irow{\sigma_{11}&\sigma_{22}&\sigma_{33}&\sigma_{23}&\sigma_{13}&\sigma_{12}}\equiv\irow{\sigma_1&\sigma_2&\sigma_3&\sigma_4&\sigma_5&\sigma_6}$ and strain $\bm{\overrightarrow{\epsilon}}=\irow{\epsilon_{11}&\epsilon_{22}&\epsilon_{33}&\epsilon_{23}&\epsilon_{13}&\epsilon_{12}}\equiv\irow{\epsilon_1&\epsilon_2&\epsilon_3&\epsilon_4&\epsilon_5&\epsilon_6}$ combined with $\bm{\overleftrightarrow{C}}=C_{ij}$ denoting either two stiffnesses $C_{11}$ and $C_{44}$ describe isotropic rocks as

\nomenclature{$\sigma_{ij}$}{\citeauthor{einstein1916} notation for stress.}
\nomenclature{$\epsilon_{kl}$}{\citeauthor{einstein1916} notation for strain.}
\nomenclature{$C_{ijkl}$}{\citeauthor{einstein1916} notation for stiffnesses.}
\nomenclature{$C_{ij}$}{\citeauthor{voigt1910} notation for stiffnesses.}
\nomenclature{$\bm{\overleftrightarrow{C}}$}{Stiffness tensor.}
\nomenclature{$\bm{\overrightarrow{\sigma}}$}{Stress vector.}
\nomenclature{$\bm{\overrightarrow{\epsilon}}$}{Strain vector.}

\begin{equation}
    \underbrace{\begin{pmatrix}
        \sigma_{11}                                                                                 \\
        \sigma_{22}                                                                                 \\
        \sigma_{33}                                                                                 \\
        \sigma_{23}                                                                                 \\
        \sigma_{13}                                                                                 \\
        \sigma_{12}
    \end{pmatrix}}_{\bm{\overrightarrow{\sigma}}}
    =
    \underbrace{\begin{pmatrix}
        C_{11}          & C_{11}-2C_{44}    & C_{11}-2C_{44}    & 0         & 0         & 0         \\
        C_{11}-2C_{44}  & C_{11}            & C_{11}-2C_{44}    & 0         & 0         & 0         \\
        C_{11}-2C_{44}  & C_{11}-2C_{44}    & C_{11}            & 0         & 0         & 0         \\
        0               & 0                 & 0                 & C_{44}    & 0         & 0         \\
        0               & 0                 & 0                 & 0         & C_{44}    & 0         \\
        0               & 0                 & 0                 & 0         & 0         & C_{44}
    \end{pmatrix}}_{\bm{\overleftrightarrow{C}}}
    \underbrace{\begin{pmatrix}
        \epsilon_{11}                                                                               \\
        \epsilon_{22}                                                                               \\
        \epsilon_{33}                                                                               \\
        2\epsilon_{23}                                                                              \\
        2\epsilon_{13}                                                                              \\
        2\epsilon_{12}
    \end{pmatrix}}_{\bm{\overrightarrow{\epsilon}}},
\label{eq:iso_cij_matrix}
\end{equation}

\nomenclature{$\sigma_{11}$}{One of the stresses $\sigma_{ij}$.}
\nomenclature{$\sigma_{22}$}{One of the stresses $\sigma_{ij}$.}
\nomenclature{$\sigma_{33}$}{One of the stresses $\sigma_{ij}$.}
\nomenclature{$\sigma_{23}$}{One of the stresses $\sigma_{ij}$.}
\nomenclature{$\sigma_{13}$}{One of the stresses $\sigma_{ij}$.}
\nomenclature{$\sigma_{12}$}{One of the stresses $\sigma_{ij}$.}
\nomenclature{$C_{11}$}{One of five stiffnesses assuming \acrshort{ti} symmetry.}
\nomenclature{$C_{44}$}{One of five stiffnesses assuming \acrshort{ti} symmetry.}
\nomenclature{$\epsilon_{11}$}{One of the strains $\epsilon_{ij}$.}
\nomenclature{$\epsilon_{22}$}{One of the strains $\epsilon_{ij}$.}
\nomenclature{$\epsilon_{33}$}{One of the strains $\epsilon_{ij}$.}
\nomenclature{$\epsilon_{23}$}{One of the strains $\epsilon_{ij}$.}
\nomenclature{$\epsilon_{13}$}{One of the strains $\epsilon_{ij}$.}
\nomenclature{$\epsilon_{12}$}{One of the strains $\epsilon_{ij}$.}

\noindent or three additional stiffnesses $C_{13}$, $C_{33}$, and $C_{66}$ describe anisotropic rocks as

\begin{equation}
    \underbrace{\begin{pmatrix}
        \sigma_{11}                                                                         \\
        \sigma_{22}                                                                         \\
        \sigma_{33}                                                                         \\
        \sigma_{23}                                                                         \\
        \sigma_{13}                                                                         \\
        \sigma_{12}
    \end{pmatrix}}_{\bm{\overrightarrow{\sigma}}}
    =
    \underbrace{\begin{pmatrix}
        C_{11}          & C_{11}-2C_{66}    & C_{13}    & 0         & 0         & 0         \\
        C_{11}-2C_{66}  & C_{11}            & C_{13}    & 0         & 0         & 0         \\
        C_{13}          & C_{13}            & C_{33}    & 0         & 0         & 0         \\
        0               & 0                 & 0         & C_{44}    & 0         & 0         \\
        0               & 0                 & 0         & 0         & C_{44}    & 0         \\
        0               & 0                 & 0         & 0         & 0         & C_{66}
    \end{pmatrix}}_{\bm{\overleftrightarrow{C}}}
    \underbrace{\begin{pmatrix}
        \epsilon_{11}                                                                       \\
        \epsilon_{22}                                                                       \\
        \epsilon_{33}                                                                       \\
        2\epsilon_{23}                                                                      \\
        2\epsilon_{13}                                                                      \\
        2\epsilon_{12}
    \end{pmatrix}}_{\bm{\overrightarrow{\epsilon}}}
    ,
\label{eq:aniso_cij_matrix}
\end{equation}

\nomenclature{$C_{13}$}{One of five stiffnesses assuming \acrshort{ti} symmetry.}
\nomenclature{$C_{33}$}{One of five stiffnesses assuming \acrshort{ti} symmetry.}
\nomenclature{$C_{66}$}{One of five stiffnesses assuming \acrshort{ti} symmetry.}

\noindent where $C_{11}$, $C_{13}$, $C_{33}$, $C_{44}$, and $C_{66}$ combinations yield the anisotropic versions of $E$, $\nu$, $G$, $K$, and $H$. For example, without further elaboration, isotropic $E$ and $\nu$ expand to anisotropic $E_\mathrm{V}$, $E_{45}$, and $E_\mathrm{H}$ as well as $\nu_\mathrm{VH}$, $\nu_\mathrm{HV}$, and $\nu_\mathrm{HH}$ for the longitudinal mode, while $H=C_{33}$ in Equation~\ref{eq:direct_p} for the uniaxial mode of a $0\degree$ specimen.

\nomenclature{$E_\mathrm{V}$}{One of three \citeauthor{young1807}'s moduli assuming \acrshort{ti} symmetry.}
\nomenclature{$E_{45}$}{One of three \citeauthor{young1807}'s moduli assuming \acrshort{ti} symmetry.}
\nomenclature{$E_\mathrm{H}$}{One of three \citeauthor{young1807}'s moduli assuming \acrshort{ti} symmetry.}
\nomenclature{$\nu_\mathrm{VH}$}{One of three \citeauthor{poisson1827}'s ratios assuming \acrshort{ti} symmetry.}
\nomenclature{$\nu_\mathrm{HV}$}{One of three \citeauthor{poisson1827}'s ratios assuming \acrshort{ti} symmetry.}
\nomenclature{$\nu_\mathrm{HH}$}{One of three \citeauthor{poisson1827}'s ratios assuming \acrshort{ti} symmetry.}

\begin{sidewaystable}
\centering

\caption{Tabulated relationships among elastic constants in an isotropic material modified from \citet{mavko2020} who based it on \citet{birch1961}. Similar tabulated versions of these relations are common in literature (even \citet{gassmann1951} included one). Given any combination of two parameters, any other elastic properties are uniquely determined from these formulae. The last row features the relationship as functions of velocities $V_\mathrm{P}$ and $V_\mathrm{S}$ as well as density $\rho$. The last column features $V_\mathrm{P}/V_\mathrm{S}$ described by different pairs of elastic constants. }

\label{tab:elastic_relations}

\begin{threeparttable}
    \begin{tabular}{cccccccc}
    \toprule[1.5pt]
    Parameters                              & $K$                                                       & $E$                                                                                                       & $\lambda_\mathrm{L}$                                                   & $\nu$                                                                     & $H$                                                                  & $\mu_\mathrm{L}$                                      & $V_\mathrm{P}/V_\mathrm{S}$                                                               \\
    \midrule
    $(\lambda_\mathrm{L},\mu_\mathrm{L})$   & $\lambda_\mathrm{L}+\frac{2}{3}\mu_\mathrm{L}$            & $\mu_\mathrm{L}\Big(\frac{3\lambda_\mathrm{L}+2\mu_\mathrm{L}}{\lambda_\mathrm{L}+\mu_\mathrm{L}}\Big)$   &                                                                       & $\frac{\lambda_\mathrm{L}}{2(\lambda_\mathrm{L}+\mu_\mathrm{L})}$         & $\lambda_\mathrm{L}+2\mu_\mathrm{L}$                                  &                                                       & $\Big(\frac{\lambda_\mathrm{L}+2\mu_\mathrm{L}}{\mu_\mathrm{L}}\Big)^{\frac{1}{2}}$       \\
    $(K,\lambda_\mathrm{L})$                &                                                           & $9K\Big(\frac{K-\lambda_\mathrm{L}}{3K-\lambda_\mathrm{L}})$                                              &                                                                       & $\frac{\lambda_\mathrm{L}}{3K-\lambda_\mathrm{L}}$                        & $3K-2\lambda_\mathrm{L}$                                              & $\frac{3(K-\lambda_\mathrm{L})}{2}$                   & $\Big(\frac{\frac{4}{3}\lambda_\mathrm{L}-2K}{\lambda_\mathrm{L}-K}\Big)^{\frac{1}{2}}$   \\
    $(K,\mu_\mathrm{L})$                    &                                                           & $\frac{9K\mu_\mathrm{L}}{3K+\mu_\mathrm{L}}$                                                              & $K-\frac{2}{3}\mu_\mathrm{L}$                                         & $\frac{3K-2\mu_\mathrm{L}}{2(3K+\mu_\mathrm{L})}$                         & $K+\frac{4}{3}\mu_\mathrm{L}$                                         &                                                       & $\Big(\frac{K+\frac{4}{3}\mu_\mathrm{L}}{\mu_\mathrm{L}}\Big)^{\frac{1}{2}}$              \\
    $(E,\mu_\mathrm{L})$                    & $\frac{E\mu_\mathrm{L}}{3(3\mu_\mathrm{L}-E)}$            &                                                                                                           & $\mu_\mathrm{L}\Big(\frac{E-2\mu_\mathrm{L}}{3\mu_\mathrm{L}-E}\Big)$ & $\frac{E}{2\mu_\mathrm{L}}-1$                                             & $\mu_\mathrm{L}\Big(\frac{4\mu_\mathrm{L}-E}{3\mu_\mathrm{L}-E}\Big)$ &                                                       & $\Big(\frac{E-4\mu_\mathrm{L}}{E-3\mu_\mathrm{L}}\Big)^{\frac{1}{2}}$                     \\
    $(K,E)$                                 &                                                           &                                                                                                           & $3K\Big(\frac{3K-E}{9K-E}\Big)$                                       & $\frac{3K-E}{6K}$                                                         & $3K\Big(\frac{3K+E}{9K-E}\Big)$                                       & $\frac{3KE}{9K-E}$                                    & $\Big(\frac{E+3K}{E}\Big)^{\frac{1}{2}}$                                                  \\
    $(\lambda_\mathrm{L},\nu)$              & $\lambda_\mathrm{L}\Big(\frac{1+\nu}{3\nu}\Big)$          & $\lambda_\mathrm{L}\frac{(1+\nu)(1-2\nu)}{\nu}$                                                           &                                                                       &                                                                           & $\lambda_\mathrm{L}\Big(\frac{1-\nu}{\nu}\Big)$                       & $\lambda_\mathrm{L}\Big(\frac{1-2\nu}{2\nu}\Big)$     & $\Big[\frac{2(1-\nu)}{(1-2\nu)}\Big]^{\frac{1}{2}}$                                       \\
    $(\mu_\mathrm{L},\nu)$                  & $\nu\Big[\frac{2(1+\nu)}{3(1-2\nu)}\Big]$                 & $2\mu_\mathrm{L}(1+\nu)$                                                                                  & $\mu_\mathrm{L}\Big(\frac{2\nu}{1-2\nu}\Big)$                         &                                                                           & $\mu_\mathrm{L}\Big(\frac{2-2\nu}{1-2\nu}\Big)$                       &                                                       & $\Big[\frac{2(1-\nu)}{(1-2\nu)}\Big]^{\frac{1}{2}}$                                       \\ 
    $(K,\nu)$                               &                                                           & $3K(1-2\nu)$                                                                                              & $3K\Big(\frac{\nu}{1+\nu}\Big)$                                       &                                                                           & $3K\Big(\frac{1-\nu}{1+\nu}\Big)$                                     & $3K\Big(\frac{1-2\nu}{2+2\nu}\Big)$                   & $\Big[\frac{2(1-\nu)}{(1-2\nu)}\Big]^{\frac{1}{2}}$                                       \\
    $(E,\nu)$                               & $\frac{E}{3(1-2\nu)}$                                     &                                                                                                           & $\frac{E\nu}{(1+\nu)(1-2\nu)}$                                        &                                                                           & $\frac{E(1-\nu)}{(1+\nu)(1-2\nu)}$                                    & $\frac{E}{2(1+\nu)}$                                  & $\Big[\frac{2(1-\nu)}{(1-2\nu)}\Big]^{\frac{1}{2}}$                                       \\
    $(H,\nu)$                               & $H-\frac{4}{3}\mu_\mathrm{L}$                             & $\mu_\mathrm{L}\Big(\frac{3H-4\mu_\mathrm{L}}{H-\mu_\mathrm{L}}\Big)$                                     & $H-2\mu_\mathrm{L}$                                                   & $\frac{H-2\mu_\mathrm{L}}{2(M-\mu_\mathrm{L})}$                           &                                                                       &                                                       & $\Big(\frac{H}{\mu_\mathrm{L}}\Big)^{\frac{1}{2}}$                                        \\
    $(V_\mathrm{P},V_\mathrm{S})$           & $\rho\Big(V_\mathrm{P}^2-\frac{4}{3}V_\mathrm{S}^2\Big)$  & $\rho V_\mathrm{S}^2\Big(\frac{3V_\mathrm{P}^2-4V_\mathrm{S}^2}{V_\mathrm{P}^2-V_\mathrm{S}^2}\Big)$      & $\rho(V_\mathrm{P}^2-2V_\mathrm{S}^2)$                                & $\frac{V_\mathrm{P}^2-2V_\mathrm{S}^2}{2(V_\mathrm{P}^2-V_\mathrm{S}^2)}$ & $\rho V_\mathrm{P}^2$                                                 & $\rho V_\mathrm{S}^2$                                 &                                                                                           \\
    \bottomrule[1.5pt]
    \end{tabular}%
\end{threeparttable}
\end{sidewaystable}

\subsection{ATTENUATION EQUATIONS}

Frequency-dependent rock behaviour caused by wave-induced fluid motions at different scales within the porous network is related to the causal-consistent dispersion and attenuation phenomena. The quality factor $Q$ is a measure of a material's dissipativity \citep{mavko2020}. \say{The greater the value of $Q$, the smaller the internal friction} \citep{birch1938} which is intuitive since, vice versa, low $Q$-values imply large dissipations. $Q$ or $Q^{-1}$ customarily describes attenuation and comes in many forms including as the imaginary $M_\Im$ and real $M_\Re$ parts of complex modulus $M$ \citep{knopoff1964}

\nomenclature{$Q$}{Quality factor.}
\nomenclature{$M$}{Complex modulus.}
\nomenclature{$M_\Im$}{Imaginary part of complex modulus $M$.}
\nomenclature{$M_\Re$}{Real part of complex modulus $M$.}

\begin{equation}
    Q^{-1}=\frac{M_\Im}{M_\Re} \label{eq:q_mi_mr}
\end{equation}

\nomenclature{$Q^{-1}$}{Inverse quality factor.}

\noindent as the ratio of energy stored $W_\mathrm{m}$ to energy loss per cycle $\Delta W$ \citep{oconnell1978}

\begin{equation}
    Q^{-1} = \frac{\Delta W}{2 \pi W_\mathrm{m}},
    \label{eq:c1-energy_attenuation}
\end{equation}

\nomenclature{$W_\mathrm{m}$}{Energy stored.}
\nomenclature{$\Delta W$}{Energy loss per cycle.}
\nomenclature{$\pi$}{Pi. \nomunit{$3.141592653\dots$}}

\noindent or as the tangent of the phase angle $\Delta\theta$ between the applied stress $\theta_{\upsigma}$ and the resulting strain $\theta_{\upepsilon}$ \citep{spencer1981}

\begin{equation}
    Q^{-1}=\tan(\underbrace{\overbrace{\theta_\mathrm{STD}}^{\theta_\upsigma}-\overbrace{\theta_\mathrm{SPE}}^{\theta_\upepsilon}}_{\Delta\theta}),
\label{eq:phase_shift}
\end{equation}

\begin{figure}[H]
    \centering
    \includegraphics[width=0.5\textwidth]{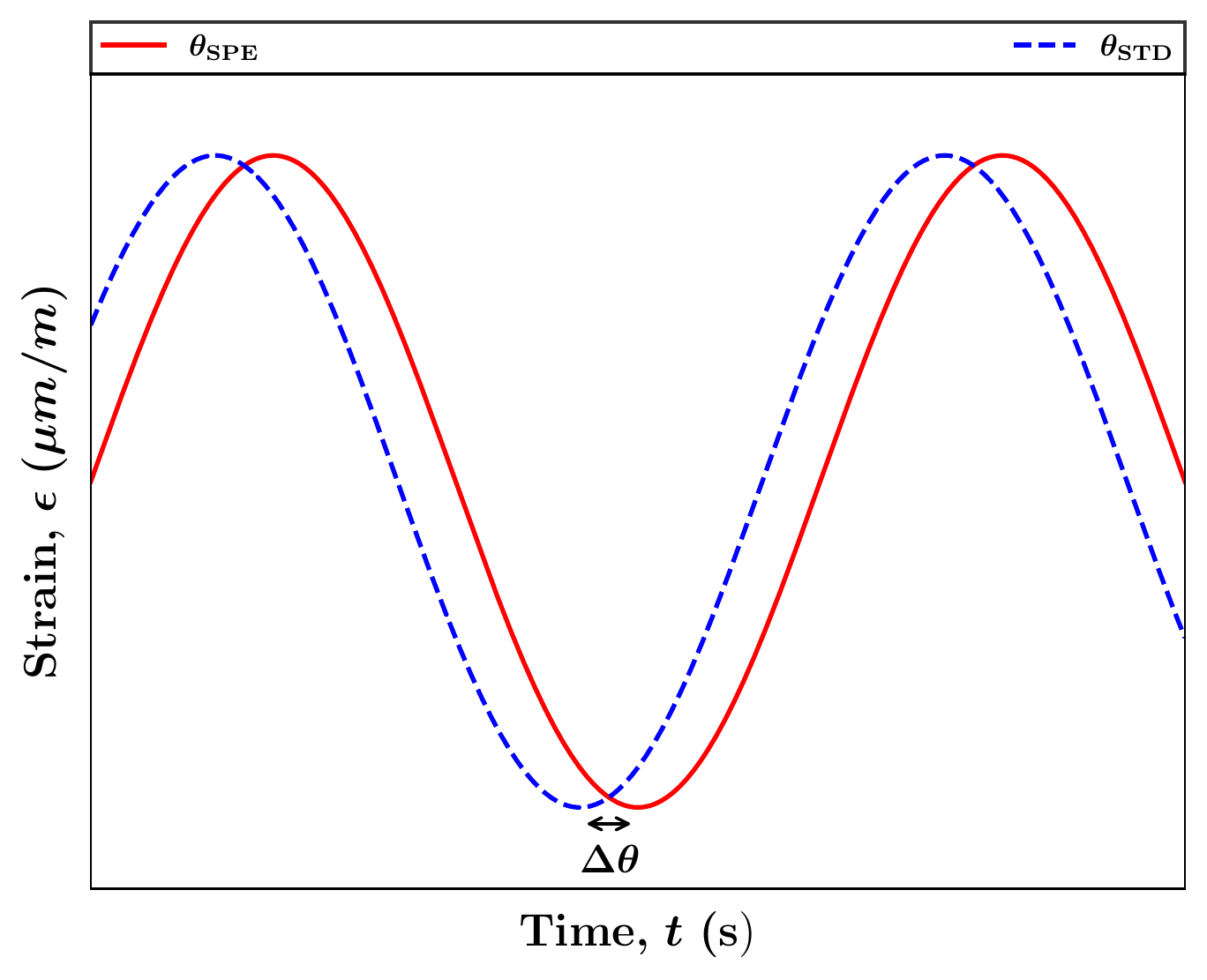}
    \caption{Equation~\ref{eq:phase_shift} graphically displayed in which the phase shift $\Delta\theta$ between $\theta_\mathrm{STD}$ and $\theta_\mathrm{SPE}$ yields $Q^{-1}$. Both strain amplitudes are equal $B_\mathrm{STD}=B_\mathrm{SPE}$ for the sake of simplicity.}
    \label{fig:phase_shift}
\end{figure}

\nomenclature{$\Delta \theta$}{Phase angle.}
\nomenclature{$\theta_\upsigma$}{Stress phase.}
\nomenclature{$\theta_\upepsilon$}{Strain phase}
\nomenclature{$\theta_\mathrm{STD}$}{Standard phase.}
\nomenclature{$\theta_\mathrm{SPE}$}{Specimen phase.}

\noindent which in practice corresponds to the phase angle $\Delta\theta$ between the standard $\theta_\mathrm{STD}$ and the specimen $\theta_\mathrm{SPE}$ exemplified by Figure~\ref{fig:phase_shift}. Equation~\ref{eq:phase_shift} is fundamental for \acrshort{fo} because oscillating stress $\sigma(t)=\sigma \mathrm{e}^{i\omega t}$ and strain $\epsilon(t)=\epsilon \mathrm{e}^{i\omega t - \Delta\theta}$ are causality-coupled, whereas \acrshort{rb} and \acrshort{pt} respectively rely on the width of the resonant peak or the time constant of the resonant decay and the \acrfull{sr} method to determine $Q$. \citet{morozov2019} questioned the usefulness of $Q$ as a measure and if its many definitions are \say{true}, \say{assumed}, or \say{apparent}.

\nomenclature{$\omega$}{Angular frequency defined as $\omega=2\pi f$.}

To evaluate mechanisms, \citet{winkler1979} assumed an isotropic solid described by complex moduli with small imaginary parts to transform between two moduli

\begin{subequations}
    \begin{align}
        \frac{(1-\nu)(1-2\nu)}{Q_\mathrm{P}}&=\frac{1+\nu}{Q_\mathrm{E}}-\frac{2\nu(2-\nu)}{Q_\mathrm{G}},         \\
        \frac{1-2\nu}{Q_\mathrm{K}}&=\frac{3}{Q_\mathrm{E}}-\frac{2(\nu+1)}{Q_\mathrm{G}},                         \\
        \frac{1+\nu}{Q_\mathrm{K}}&=\frac{3(1-\nu)}{Q_\mathrm{P}}-\frac{2(1-2\nu)}{Q_\mathrm{G}},
    \end{align}
\end{subequations}

\noindent with $Q_\mathrm{E}=E_\Re/E_\Im$, $Q_\mathrm{K}=K_\Re/K_\Im$, and $Q_\mathrm{G}=G_\Re/G_\Im$ while  simultaneously proving that one of these attenuation relations must be true

\begin{subequations}
    \begin{align}
        Q_\mathrm{K}>Q_\mathrm{P}&>Q_\mathrm{E}>Q_\mathrm{G}, \label{eq:true1}               \\
        Q_\mathrm{K}<Q_\mathrm{P}&<Q_\mathrm{E}<Q_\mathrm{G}, \label{eq:true2}               \\
        Q_\mathrm{K}=Q_\mathrm{P}&=Q_\mathrm{E}=Q_\mathrm{G}.
    \end{align}
\end{subequations}

\nomenclature[B]{$Q_\mathrm{E}$}{Quality-factor of \citeauthor{young1807}'s modulus $E$.}
\nomenclature[B]{$Q_\mathrm{K}$}{Quality-factor of bulk modulus $K$.}
\nomenclature[B]{$Q_\mathrm{G}$}{Quality-factor of shear modulus $G$.}
\nomenclature[B]{$E_\Im$}{Imaginary part of \citeauthor{young1807}'s modulus $E$.}
\nomenclature[B]{$E_\Re$}{Real part \citeauthor{young1807}'s modulus $E$.}
\nomenclature[B]{$K_\Im$}{Imaginary part of bulk modulus $K$.}
\nomenclature[B]{$K_\Re$}{Real part of bulk modulus $K$.}
\nomenclature[B]{$G_\Im$}{Imaginary part of shear modulus $G$.}
\nomenclature[B]{$G_\Re$}{Real part of shear modulus $G$.}

\citet{johnston1979} summarized a series of $Q$-dependencies based on the accumulation of individual attenuation measurements (frequency, strain amplitude, fluid saturation, pressure and stress, and temperature) and elaborated on attenuation mechanisms: (i) matrix anelasticity, (ii) viscosity and flow of saturating fluids, and (iii) scattering from inclusions. Mechanism (i) is related to solid friction between grains in dry rocks. Mechanism (ii) includes local \citeauthor{biot1956a} flow within the pore space often termed \say{squirt-flow} \citep{biot1956a}, flow in heterogeneous media \citep{white1975}, and flow related to the wavelength-scale equilibration \citep{biot1962} at the respective microscopic, mesoscopic, and macroscopic scales. Mechanism (iii) tends to be significant for wavelengths close to the size of any heterogeneities. Mechanism (ii) is typically caused by \acrfull{wiff} \citep{muller2010} but evidence for \acrfull{wiged} \citep{tisato2015} also exists. Fundamental to poroelasticity \citep{gassmann1951,biot1956a,biot1956b,rice1976} are the opposing concepts of (i) drained and (ii) undrained conditions defined as the fluid's ability (i) or inability (ii) to flow in or out of the porous medium. \citet{biot1962} predicted the existence of two regimes of fluid motion governed by (i) viscous effects at the low frequency limit \citep{biot1956a} and by (ii) inertia effect at the high frequency limit \citep{biot1956b}. Equal viscous and inertial forces of the fluid occur at \citeauthor{biot1962}'s characteristic frequency $f_\mathrm{b}$ while viscous flow and inertial drag respectively dominate below and above. These concepts define characteristic frequencies for the drained-undrained and quasistatic-dynamic transitions. Low- and high-frequency behaviours (\say{relaxed} versus \say{unrelaxed} states) are distinguished by characteristic frequencies $f_\mathrm{c}$ at different scales: microscopic $f_\mathrm{c} \rightarrow f_\mathrm{s}$, mesoscopic $f_\mathrm{c} \rightarrow f_\mathrm{p}$, and macroscopic $f_\mathrm{c} \rightarrow f_\mathrm{d},f_\mathrm{b}$ are respectively defined as 

\nomenclature{$f_\mathrm{c}$}{Characteristic frequency.}

\begin{align}
    f_\mathrm{s}&=\frac{\xi^3K_\mathrm{d}}{\eta},\\
    f_\mathrm{p}&=\frac{\kappa K_\mathrm{S}}{\pi L^2 \eta},\\
    f_\mathrm{d}&=\frac{\kappa}{\eta}\frac{K_\mathrm{u}}{L^2},\\
    f_\mathrm{b}&=\frac{\eta\phi}{2\pi\rho_\mathrm{f}\kappa},
\end{align}

\nomenclature{$f_\mathrm{s}$}{\say{Squirt-flow} characteristic frequency.}
\nomenclature{$\xi$}{Aspect ratio.}
\nomenclature{$K_\mathrm{d}$}{Drained bulk modulus.}
\nomenclature{$\eta$}{Viscosity.}

\nomenclature{$f_\mathrm{p}$}{Patchy-saturation characteristic frequency.}
\nomenclature{$\kappa$}{Permeability.}
\nomenclature{$K_\mathrm{S}$}{Solid bulk modulus.}

\nomenclature{$f_\mathrm{d}$}{Drained-undrained characteristic frequency.}
\nomenclature{$K_\mathrm{u}$}{Undrained bulk modulus.}

\nomenclature{$f_\mathrm{b}$}{\citeauthor{biot1962}'s characteristic frequency.}
\nomenclature{$\phi$}{Porosity.}
\nomenclature{$\rho_\mathrm{f}$}{Pore fluid density.}

\noindent where $\xi$ is aspect ratio, $\eta$ is viscosity, $\kappa$ is permeability, $\phi$ is porosity, $\rho_\mathrm{f}$ is pore fluid density, while $K_\mathrm{d}$, $K_\mathrm{u}$, and $K_\mathrm{S}$ are drained, undrained, and solid bulk moduli \citep{delle_piane_2014,mavko2020}.

\nomenclature{$B_\mathrm{STD}$}{Strain amplitude of standard.}
\nomenclature{$B_\mathrm{SPE}$}{Strain amplitude of specimen.}

Bulk and shear viscosities are dissimilar but the controlling molecular motion are probably similar \citep{hasanov2016}. The former is not only poorly defined but also rarely measured compared to the latter. Viscosity discernibility is crucial since bulk dominates shear in dispersivity and attenuativity of rocks saturated with highly viscous fluids. Moreover, stronger bulk losses characterize saturated sediments comprising rounded sand grains rather than silt-sized quartz particles \citep{prasad1992}.

\subsection{MODELS}

\say{All models are wrong but some are useful}\footnote{Common aphorism generally attributed to \citet{box1976}.} in simulating experimental behaviour and isolating the dominant mechanisms. \say{Numerous theories on wave propagation exist but these concepts remain unencumbered by measured data needed to prove, delimit or extend them} \citep{batzle2006}. An increasing number of phenomenological and non-phenomenological (predictive) models that combine flow at different scales are continuously being proposed. Since different experimental techniques cover different frequencies, models that simulate dispersive behaviour across frequencies are imperative to unite results from different experiments. \acrfull{cc} \citep{colecole1941} and \acrfull{sls} (or \citeauthor{zener1948}) \citep{zener1948} models are among the most common ones alongside the analytical \acrfull{kkr} \citep{kronig1926,kramers1927}. \acrshort{cc} is based on \acrshort{kkr} and reduces to the fundamental \citeauthor{debye1929} model \citep{debye1929} if $\alpha\in[0,1] \rightarrow \alpha=0$, while \acrshort{sls} is based on the \citeauthor{maxwell1867} \citep{maxwell1867} and \citeauthor{kelvin1865}-\citeauthor{voigt1892} \citep{kelvin1865,voigt1892} models but with different combinations of springs and dashpots to represent elastic and viscous components, respectively. \citeauthor{burgers1935} viscoelastic model \citep{burgers1935} is a combination of the other models (\citeauthor{maxwell1867}, \citeauthor{kelvin1865}-\citeauthor{voigt1892}, and \citeauthor{zener1948}). \acrshort{kkr} are the bidirectional mathematical solutions to the problem of connecting the real and imaginary parts of any complex function that is constrained by causality and analytic in the upper half-plane. \citet{spencer1981} extended the frequency-dependent \acrshort{cc} model from the realm of the dielectric constant of liquids to the realm of the real and imaginary parts of porous media moduli impacted by relaxation processes. \citet{mikhaltsevitch2016b} considered an isotropic viscoelastic material regarded as a dynamic system for which the stress-strain relationship is linear with given mechanical properties as a numerical solution of \acrshort{kkr}. In general, \acrshort{kkr} is superior to \acrshort{cc} due to its analytical nature but necessitates mechanical and attenuation measurements at all frequencies unlike \acrshort{cc} that can be fitted using only mechanical measurements. \citet{rorheim2022phd} describes a joint-fit-based procedure based on \acrshort{cc} that uses both mechanical and attenuation measurements as input. \acrshort{cc} is however limited to describing one relaxation time, transition, or mechanism albeit multiple may concurrently exist.

\nomenclature{$\alpha$}{\acrlong{cc}'s spreading factor. \nomunit{$\in[0,1]$}}


\section{METHODS} \label{sec:methods}

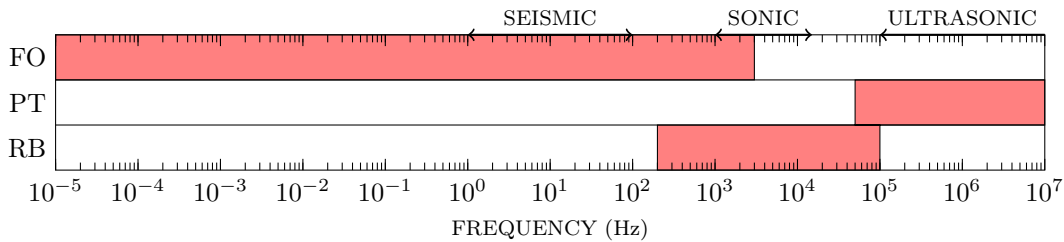
\begin{figure}[H]
\centering
\begin{tikzpicture}
    \pgfplotstableread{
        Label Start  Stop
        1     2e2    1e5 
        2     5e4    1e7
        3     1e-5   3e3 
    }\datatable
    \pgfplotsset{
      every axis/.style={
        width=1\textwidth,
        y=0.6cm,
        bar width=0.6cm,
        enlarge y limits=0.25,
        label style={font=\footnotesize},
        axis on top,
        xbar stacked,
        xmin=1e-5, xmax=1e7,
        xmode=log, 
        ytick={1,...,3},
        yticklabels={\acrshort{rb}, \acrshort{pt}, \acrshort{fo}},
        ytick style={draw=none},
        extra y ticks={1.5,2.5},
        extra y tick labels={},
        extra y tick style={grid=minor},
        grid style={black},
        every tick/.style={black},
        xlabel={FREQUENCY (Hz)},
        clip=false,
        stack negative=on previous,
      },
      minimum/.style={forget plot, draw=none, fill=none},
    }
  \begin{axis}
    \draw [<->, thick] (axis cs:1,3.5) -- (axis cs:1e2,3.5) node [midway,above, font=\footnotesize] {SEISMIC};
    \draw [<->, thick] (axis cs:1e3,3.5) -- (axis cs:1.5e4,3.5) node [midway,above, font=\footnotesize] {SONIC};
    \draw [<-, thick] (axis cs:1e5,3.5) -- (axis cs:1e7,3.5) node [midway,above, font=\footnotesize] {ULTRASONIC};
    \addplot [minimum] table [x=Start, y=Label] {\datatable};
    \addplot[black, fill=red!50] table [y=Label, x expr=\thisrowno{2}-\thisrowno{1}] {\datatable};
  \end{axis}
\end{tikzpicture}
\caption{Probeable frequencies subdivided into seismic, sonic, and ultrasonic for the three primary techniques: \acrshort{rb}, \acrshort{pt}, and \acrshort{fo}.}
\label{fig:freq_range}
\end{figure}

Experimental determination of dynamic rock properties chronologically began with \acrfull{rb}, \acrfull{pt}, and \acrfull{fo} at the respective sonic, ultrasonic, and seismic frequencies (Figure~\ref{fig:freq_range}). All three techniques are elaborated but \acrshort{fo} in greatest detail with (to our knowledge) all accessible studies described and summarized.

\subsection{RESONANT BAR (RB)}

\citet{ide1935} invented a technique based on \citet{cady1922,quimby1925,boyle1931} for dynamic determination of rock properties that produced longitudinal vibrations of rods by electrostatic traction. Inspired by the novelty of their colleague's approach, \citet{birch1938} studied the longitudinal, flexural, and torsional modes of a long granite column forced into resonant vibrations excited by an alternating magnetic field. This ultimately became known as the widely acknowledged \acrfull{rb} technique whose function is to probe sonic frequencies. However rare measurements in the sonic frequency range may be, experiments using long and slender samples are less so than experiments involving conventional core specimens. The longer the resonator, the lower the characteristic frequency. Since frequency is lowered as a function of bar length, it is common practice to cement short-length bars together due to the difficulty to directly core sufficiently long bars \citep{born1941}. A sinusoidal force is traditionally submitted to one end of a rock bar fixed at its center of mass (which also coincides with the location of minimum displacement) while the resultant vibrations are measured at the other. Piezoelectric or electromagnetic transducers excite either the extensional or the torsional mode of vibration. Moduli are calculated from the resonance frequencies that depend on the material’s velocity, dimensions, and density. Attenuation is determined by the width of the resonant peak or the time constant of the resonant decay.

Despite the obvious advantage of probing sonic frequencies, \citet{wang1997} identified (i) extensive system calibration and data corrections as well as difficulty with (ii) obtaining high-pressure, high-temperature data and (iii) pure shear mode as disadvantages. Disadvantageous is also that the characteristic frequency is a function of the bar length. Although \citet{jones1983} extended the investigations of the effects of confining pressure, pore pressure, and degree of saturation by \citet{winkler1982} with elevated temperatures, \acrshort{rb} measurements at \textit{in-situ} conditions belong to the rarities. \acrshort{rb} suffered dimensional limitations until \citet{tittmann1977} proposed the possibility to reduce resonance frequencies by the use of additional mass that changes the moment of inertia without altering the rigidity. These words resonated with \citet{nakagawa2011} who realized that small specimens could be explored using a modified version of \acrshort{rb}. \citet{nakagawa2011} designed an apparatus geometrically similar to the Split-\citeauthor{hopkinson1914} Pressure Bar (\acrshort{shpb}) described by \citet{kolsky1949} and named after \citet{hopkinson1914}. \citet{nakagawa2011} added mass to his \acrfull{shrb} by a pair of $40.6$ cm steel extension bars with attached \acrshort{pzt} source and accelerometer receivers. Extensive corrections for the jacket and the interfaces are however applied in a numerical inversion model to correct for the specimen-rod interface effects. Table~\ref{tab:review} lists authors and techniques that have successfully probed the sonic frequency range as functions of frequency and specimen dimensions categorized into measured parameters. Featured is not only \acrshort{rb} and \acrshort{fo} but also the \acrfull{ptu}, \acrfull{ghrc}, and \acrfull{rus} techniques.

\subsection{PULSE-TRANSMISSION (PT)}

\acrfull{pt} was extended from metals \citep{hughes1949} to rocks \citep{hughes1950,hughes1951,wyllie1956} and further disseminated \citep{birch1960,toksozetal1979,winkler1983} as a technique. \acrshort{pt} determines the travel-time of an elastic wave propagating between two \acrshort{pzc} (normally quartz or \acrshort{pzt}) transducer elements in which one generates and the other records the ultrasonic pulse. The piezoceramic size determines its resonant (or center) frequency. Scattering effects are avoided if specimen heterogeneities are eclipsed in size by the wavelength. Since distance travelled (equals specimen length) and travel-time are known, the basic equation of motion yields velocity based on the time of flight principle. \acrfull{pzc} elements deform mechanically and generate electricity respectively as external voltages and forces are applied. P- and S-wave transducers are sensitive to their respective wave types. Polarization-alignment is paramount for the S-wave transducers due to the physical nature of S-waves. In practice, the quality of the signal depends on the transducer-specimen coupling being adequate or inadequate while the accuracy of the measurements is related to the adequacy of the velocity picks. Normal P-waves are trivial but oblique P-waves and any S-waves are non-trivial to pick. Practice and theory are advanced by transducer and algorithm novelty. Eternal is the group versus phase velocity conundrum for oblique P-waves that depend on transducer size. \acrshort{pt} attenuation is determined by the \acrfull{sr} method which compares spectral amplitudes at different distances from the slope of the logarithmic decrement \citep{mavko2020}. Otherwise attenuation is also determined by comparing two geometrical identical specimens: a non-dispersive reference specimen and the studied specimen \citep{bourbie1985}. This is the same principle as for \acrshort{fo} in comparing non-dispersive reference specimens with dispersive rock specimens. \acrfull{pe} is physically different from \acrshort{pt} (echo versus transmission) but principally similar in determining attenuation by standard-specimen comparisons. 

\subsection{FORCED-OSCILLATION (FO)}

\citet{bruckshawetal1961,peselnick1961,usher1962} performed the first \acrfull{fo} measurements that determined the intrinsic attenuation and elastic moduli of rocks. The two extremes investigated the longitudinal mode (\citeauthor{young1807}'s modulus $E$ and attenuation $Q_\mathrm{E}^{-1}$), whereas the middle developed a pendulum able to generate harmonic torques (shear modulus $G$ and attenuation $Q_\mathrm{G}^{-1}$). Truth be told, internal friction of metals \citep{norton1939,ke1949} and polymers \citep{fukada1951,fukada1954,koppelmann1958} were measured by \acrshort{fo} before extended to rocks \citep{bruckshawetal1961,peselnick1961,usher1962}. Strain amplitude was not controlled in the first generation of \acrshort{fo} experiments but was later constrained to $10^{-6}$ to remain within the linear elastic domain \citep{winkler1995}. \citet{spencer1981} measured the stress-strain behaviour of rocks via transducers recording the axially applied sinusoidal force and the resulting displacement. \citet{adelinet2010} applied hydrostatic oscillations to measure bulk modulus $K$ of isotropic rocks via confining pressure pulsations. \citet{jackson2011} modified their torsional capabilities to also accommodate for measurements of the flexural mode which \citet{woirgard1971,woirgard1978} also measured for metals and rocks in the past. Unlike \citet{adelinet2010}, \citet{lozovyi2019} enforced uniaxial instead of hydrostatic conditions by adding axial oscillations to investigate anisotropic rocks. \citet{suarez2001} measured uniaxial modulus decades before \citet{lozovyi2019} without disclosing any details about the apparatus. \acrshort{fo} requires (i) a force generator with alterable frequency, (ii) a force sensor to estimate the applied stress, and (iii) strain sensors for the specimen. Modulus and attenuation are given by the stress-strain ratio and the phase angle between the stress and the strain (or the area of the hysteresis loop), respectively. If an elastic standard is chosen over a transducer to measure the stress, modulus and attenuation are given by the relative strain amplitude of the standard versus the specimen and the phase angles between the standard and the specimen. Table~\ref{tab:FO} tabulates apparatuses for (i) longitudinal, (ii) torsional, (iii) flexural, (iv) volumetric, and (v) uniaxial stress-strain oscillations. Type (ii), (iii), and (iv), and (v) apparatuses exist but type (i) is still predominant.

\nomenclature{$Q_\mathrm{E}^{-1}$}{Inverse quality factor of \citeauthor{young1807}'s modulus $E$.}
\nomenclature{$Q_\mathrm{G}^{-1}$}{Inverse quality factor of shear modulus $G$.}

\acrshort{fo} is theoretically trivial but practically non-trivial. Misalignments related to manufacturing tolerances primarily affecting attenuation measurements (inconsistent phase angles) are particularly challenging \citep{liu1980,liu1983,peselnick1987}. In the case of \citet{rorheim2019}, this problem originated from misaligned interfaces. Uneven stress distribution may be observed on the semiconductor resistivities during axial loading. Numerous approaches have been applied to alleviate systematic errors caused by misalignment if all interfaces are not flawlessly parallel and smooth when manufactured. \citet{paffenholzetal1989,takei2011,ikeda2020} neutralized potential misalignments with three axially symmetrical transducers. Resonances enforce a natural upper limit for \acrshort{fo} measurements in the low-frequency regime. Due to the nature of resonances (nodes and antinodes as well as their distribution), \citet{batzle2006} extended this limit by identifying the antinodal frequencies and adjust the measurements accordingly. Fluid-flow related dispersion occurring at the intermediate sonic frequency range may be inferred by using a very viscous fluid as described by \citet{fortin2014,pimienta2016}. Since the time-scale of diffusion processes are related to the dynamic viscosity of the fluid, the frequency of the measurement can be scaled by the viscosity of the fluid \citep{borgomano2018}. 

Despite being less resonance-prone than other transducers, strain gauges are largely limited to investigating pore-scale processes \citep{chapman2018}. Averaging multiple strain measurements at different positions on the specimen surface may however approximate the bulk mechanical properties of a rock \citep{adam2009phd}. The smaller the specimen, the better the averaging result (proportional to the covered area). \citet{chapman2018} favoured measuring bulk strain (cantilevers) in addition to or instead of local strains (strain gauges) due to the locality of heterogeneities. Bulk-strain approaches are superior for isotropic rather than anisotropic rocks due to \citeauthor{poisson1827}'s ratios being problematic to measure. \citet{tisato2013phd} explained the double-interface problem related to radial cantilevers on jacketed (sealed) specimens. \citet{spencer1981} however solved this problem with non-contact capacitive sensors and \citet{tisato2013phd} explored \citep{hall1879} effect sensors. Fibre optic \acrshort{das} is another possibility \citep{adam2009phd,yurikov2021a,yurikov2021b,yurikov2022} that faces similar problems. Simultaneous bulk and local strain measurements could perhaps also distinguish between microscopic and mesoscopic dispersion and attenuation mechanisms \citep{chapman2018}. Unlike \citet{adambatz2008} but like \citet{chapman2018}, \citet{sun2020b} analyzed individual strains before averaging to distinguish local from global flow. Stress-strain signal magnitudes and phases are computed by \acrshort{fft} after being digitized and averaged by physical or virtual lock-in amplifiers.


\acrshort{fo} is applied to measure properties of all sedimentary rock types: sandstones \citep{spencer1981,paffenholzetal1989,lienertetal1990,yin1992,cherry1996,chelidze1996,batzle2001,gautam2003,batzle2003,hofmann2006,batzle2006,madonna2010,tisato2012,david2012,madonna2013,nakagawa2013LF,tisato2013,spencer2013,yao2013,batzle2014,mikhaltsevitch2014,spencer2016,riviere2016,chapman2016,ma2016,mikhaltsevitch2016b,massaad2016,tisato2016,pimienta2017,sun2017,sun2018,agofack2018,chapman2019,yin2019,chavez2019,li2020a,borgomano2020,gallagher2020,sun2020,li2020c,ogunsami2020,sun2020b,zhao2021,yurikov2021a,chapman2021,tisato2021,han2021,he2021,ma2021,yurikov2021b,yin2021,riabokon2021a,lu2021phd,chapman2022,chapman2022b,he2022,mews2022,han2022}, carbonates \citep{peselnick1961,spencer1981,paffenholzetal1989,gautam2003,hofmann2006,batzle2006,adam2006,behura2007,adambatz2008,adam2009,zhao2014,zhao2015,mikhaltsevitch2016a,borgomano2017,chavez2019,mikhaltsevitch2019a,mikhaltsevitch2019b,borgomano2019,li2020a,riabokon2021b,mikhaltsevitch2022,sun2022,gallagher2022}, and shales \citep{suarez2001,batzle2005,duranti2005,hofmann2006,sarker2010,delle_piane_2011,delle_piane_2014,batzle2014,mikhaltsevitch2015,huang2015,szewczyk2017,mikhaltsevitch2017,li2017,lozovyi2018phd,chavez2019,mikhaltsevitch2021b,xiao2021,li2022,long2022}. \acrshort{fo} is also applied to igneous rocks and other polycrystalline materials \citep{woirgard1978,peselnick1979,spencer1981,jackson1984,gueguen1989,gribb1998a,gribb1998b,adelinet2010,jackson2011,saltiel2017,fliedner2019,cao2021,fliedner2021,fliedner2022} that are beyond the scope of this study (due to their non-sedimentary nature) but are nonetheless represented in Table~\ref{tab:FO}. \citet{suarez2001} is the only inter-frequency study comparing \acrshort{fo}, \acrshort{rb}, and \acrshort{pt}. \citet{mccann2019} used \acrshort{fo} to measure fluid properties (bulk modulus $K$ and attenuation $Q_\mathrm{K}^{-1}$) at high pressures. Be advised that Table~\ref{tab:FO} may list the same apparatus multiple times because they are continuously being improved or adjusted with time. \citet{subramaniyan2014} summarized attenuation studies on conventional reservoir rocks (i.e. sandstones and carbonates), while recent studies on conventional rocks as well as unconventional shales are added herein to update and complete the list:

\nomenclature{$Q_\mathrm{K}^{-1}$}{Inverse quality factor of bulk modulus $K$.}

\begin{enumerate}[label={\textbf{Carbonates}},left=0cm, align=left]
    \item[\textbf{Sandstones}] \citet{spencer1981} initially observed negligible $Q_\mathrm{E}^{-1}$ of a vacuum-dried Navajo specimen that eventually became significant once saturated with water, ethanol, or n-decane. \citet{paffenholzetal1989} evaluated four different sandstones at different saturation levels and observed that $E$ and $G$ decrease with increasing saturation at $<50\%$ saturation but is independent of water content at $>50\%$ saturation. In comparison, $Q_\mathrm{E}^{-1}$ and $Q_\mathrm{G}^{-1}$ are less pronounced with decreasing saturation and their peaks shift towards higher frequencies. \citet{yin1992} studied Berea specimens using \acrshort{fo} and \acrshort{rb}: $Q_\mathrm{E}^{-1}$ increases with frequency and brine-saturation. \citet{cherry1996} saw $E$ stiffening and $Q_\mathrm{E}^{-1}$ weakening of a Lyon's specimen with decreasing saturation. \citet{chelidze1996} also studied Lyon specimens but instead the effect of acetone and water-surfactant on $E$ and $Q_\mathrm{E}^{-1}$. $Q_\mathrm{E}^{-1}$ of dry or fluid-saturated ($<50\%$) Berea specimens are low and approximately frequency-independent \citep{lienertetal1990,madonna2010,tisato2012,madonna2013,tisato2013,tisato2014,chapman2016} or high and frequency-dependent \citep{tisato2012,nakagawa2013LF,yao2013,mikhaltsevitch2014,madonna2013,kuteynikova2014,chapman2016}. \citet{batzle2006} attributed such behaviour to fluid mobility $M_\mathrm{F}$ defined as $M_\mathrm{F}=\kappa/\eta$ where $k$ is permeability and $\eta$ is viscosity without presenting $Q_\mathrm{E}^{-1}$ of actual rocks: deflating $M$ is coupled with inflated relaxation time required to equilibrate the pore pressure. Fluid-saturated rocks will consequently appear stiffer under conditions in which the period of elastic perturbation is shorter than the fluid relaxation time than at lower frequencies. \citet{david2012} attempted to measure $E$, $Q_\mathrm{E}^{-1}$, and $\nu$ of dry, water and glycerine-saturated Fontainebleau specimens but ultimately questioned the reliability of the results due to inconsistencies related to \acrshort{fo} eclipsing \acrshort{pt} in $E$ and $\nu$ magnitudes. \citet{tisato2013} explained their high and frequency-dependent $Q_\mathrm{E}$ measurements for a nearly fully water-saturated Berea specimen by \acrshort{wiff} at the mesoscopic scale. \citet{mikhaltsevitch2014} observed decreasing and frequency-dependent $Q_\mathrm{E}^{-1}$ with increasing confining pressure culminating at $1$ Hz for a fully water-saturated Donnybrook specimen. \citet{spencer2013} measured $E$ and $Q_\mathrm{E}^{-1}$ as well as $\nu$ and $Q_\upnu^{-1}$ in McMurray bitumen sand specimens containing residual air where calculated $Q_\mathrm{P}^{-1}$, $Q_\mathrm{K}^{-1}$, and $Q_\mathrm{G}^{-1}$ peaks shifted towards lower frequencies as viscosity increased (by decreasing temperature) demonstrating strongly viscosity-dependent attenuation mechanisms. 
    
\nomenclature{$M_\mathrm{F}$}{Fluid mobility.}
    
    \citet{tisato2015} provided the first experimental evidence of \acrshort{wiged} in gas (air, \acrshort{n2}, or \acrshort{co2}) and water equilibrated Berea specimens which was also numerically supported. \citet{subramaniyan2015} disclosed that \say{a decrease in viscosity of the saturating fluid shifted the attenuation curve to higher frequencies} while \say{an increase in confining pressure caused a decrease in the overall magnitude of attenuation} in Fontainebleau specimens. \say{Squirt-flow} is implied as the dominant mechanism for their glycerine data. \citet{pimienta2015b,pimienta2015a} studied $E$ and $K$ frequency dependence and attenuation in water and glycerine-saturated Fontainebleau specimens. Two frequency-dependent phenomena interpreted as drained-undrained and undrained-unrelaxed $E$-transitions \citep{pimienta2015b} and underestimation of the drained-undrained transition implying a direct drained-unrelaxed $K$-transition \citep{pimienta2015a} are observed. \citet{spencer2016} concluded that \say{the modulus dispersion and attenuations ($Q_\mathrm{P}^{-1}$ and $Q_\mathrm{G}^{-1}$) in saturated sandstones are caused by a pore-scale, local-flow mechanism operating near grain contacts.} Higher harmonics allowed \citet{riviere2016} to study frequency, pressure, and strain dependence of nonlinear elasticity in dry Berea specimens. \citet{chapman2016} measured $E$ and $Q_\mathrm{E}^{-1}$ of a Berea specimen at various saturation levels and confining pressure: attenuation negligible at $<80\%$ and significant at $>91\%$ saturation also reduced and shifted towards lower frequencies by increasing confining pressure. Consistent with \acrshort{wiff} in response to heterogeneous water distribution in the pore space (patchy saturation), high enough fluid pressure to ensure full saturation also renders attenuation negligible. \citet{ma2016} attempted to validate their apparatus by aluminium and lucite measurements: $E$ and $\nu$ but no $Q_\mathrm{E}^{-1}$. $V_\mathrm{P}$ and $V_\mathrm{S}$ converted from $E$ and $\nu$ also increased with pressure and saturation. \citet{mikhaltsevitch2016b} demonstrated the causality-consistency of dry and water-saturated Donnybrook and Harvey as well as glycerol-saturated Berea specimens using \acrshort{kkr}. \citet{massaad2016, agofack2018} researched the effect of \acrshort{co2} on $E$ and $\nu$ transformed into $V_\mathrm{P}$ and $V_\mathrm{S}$ in a Berea specimen at Hz and MHz. \citet{tisato2016} proved the feasibility of combining \acrshort{fo} with $\mu$\acrshort{ct} \citep{tisato2015b} to study a dry and partially saturated Berea specimen from $0.1$ to $25$ Hz in a \acrshort{ct}-transparent cell. Insignificant changes between these different conditions aside, this was the first time elastic measurements and imaging was combined for \acrshort{fo}. Explained by a transition from \acrshort{wiff} to \acrshort{wiged} and inspired by \citet{tisato2015}, \citet{chapman2017} observed a significant steepening of the high-frequency asymptote of the measured attenuation in Berea specimens caused by a minor change in water saturation (but a significant modification in the pore fluid distribution). \citet{pimienta2017} reported drained-undrained transitions in Wilkenson, Berea, and Bentheim specimens saturated by fluids of varying viscosities. Berea features $E$ and $\nu$ dispersion as well as $Q_\mathrm{E}^{-1}$ and $Q_\upnu^{-1}$ attenuation towards higher frequencies. Bentheim and Wilkenson only feature $E$ dispersion and $Q_\mathrm{E}^{-1}$ attenuation. \citet{pimienta2017} interpreted Wilkenson's consistency and Bentheims's inconsistency with \citeauthor{zener1948}'s model \citep{zener1948} as \say{squirt-flow} or measurement inaccuracy (possibly also another physical effect). \citet{sun2017} measured two attenuation peaks probably caused by two different mechanisms for a tight sandstone saturated between $45$ and $85\%$ but no peaks beyond these limits. 
    
    \citet{sun2018} focused on presenting an enhanced \acrshort{fo} version void of subsonic resonances based on numerical modelling but also included inconclusive $E$ and $\nu$ measurements. \citet{chapman2019} observed \say{frequency-dependent attenuation and the associated moduli dispersion in response to the drained–undrained transition ($0.1$ Hz) and \say{squirt-flow} ($>3$ Hz)} in Berea specimens. \citet{yin2019} sought to elucidate \citeauthor{gassmann1951}'s fluid substitution theory by studying a clay-bearing Th{\"{u}}ringen specimen by combining \acrshort{fo} and \acrshort{pt}. Dry specimens are non-dispersive (and thus non-attenuative) whereas water-saturated specimens are dispersive (and thus attenuative) which is attributed to the drained-undrained transition. \citeauthor{gassmann1951}'s theory is consistent with their undrained $K$ but not with their water-softened $G$ which is significant at seismic but masked by \say{squirt-flow} stiffening at ultrasonic frequencies. \citet{yin2019} ascribed this reduction in surface free energy to chemical interaction between pore fluid and rock frame which eludes \citeauthor{gassmann1951}'s theory. \citet{li2020a} observed a broad distribution of $E$ and $Q_\mathrm{E}^{-1}$ across $1$-$1000$ Hz with peak attenuation at $60\%$ saturation in a tight sandstone. \citet{borgomano2020} enforced drained or undrained conditions on glycerine-saturated or -unsaturated Bleurswiller specimens for $K$ and $Q_\mathrm{K}^{-1}$ determination with microvalves (combined dead volume less than $40$ $\mu$l). \citet{gallagher2020} focused on dry, brine-, and glycerine-saturated sandstones. \citet{sun2020} compacted Bleurswiller specimens beyond \say{the critical pressure which characterizes the onset of pore collapse and grain crushing} to the point that the critical frequency of \say{squirt-flow} dispersion (relaxed-unrelaxed transition) was shifted towards higher frequencies beyond the seismic band \say{allowing \citeauthor{biot1956b}-\citeauthor{gassmann1951} to fully apply}. This result was interpreted as a consequence of increased crack aspect ratio after compaction. \citet{li2020c} complimented the tight rocks of \citet{li2020a} with a weakly consolidated sandstone. Also a study about saturation effects, peak attenuation occurs at $60$ Hz and at $79\%$ saturation. \citet{ogunsami2020} performed the first systematic inter-laboratory study applying three different \acrshort{fo} devices to cross-validate the elastic properties of a reservoir sandstone. Inter-laboratory frequency and fluid-saturation effects are consistently proven pressure-dependent at dry and decane-saturated conditions. \citet{sun2020b} investigated the impact of microstructure heterogeneity and local measurements on dispersion and attenuation of dry plus brine- and oil-saturated Shahejie specimens from $1$ to $300$ Hz. Local diverges from global flow by being influenced by the position of the strain gauges which \citet{sun2020b} attributed to crack-aspect-ratio heterogeneity since porosity and crack density are homogeneous. \citet{sun2020b} measured not only $E$ and $Q_\mathrm{E}^{-1}$ but also $\nu$ and $Q_\upnu^{-1}$ owing to biaxial semiconductor strain gauges. 
    
    \citet{zhao2021} discovered one or two attenuation peaks related to micro- and mesoscopic flow between $1$ and $2000$ Hz affected by saturation degree, oil viscosity, and confining pressure. A dual-scale fluid flow model suggested \say{that the attenuation mechanisms at different scales interplay with each other and jointly dominate the attenuation behavior of the partially saturated specimen.} \citet{yurikov2021a,yurikov2021b} explored fibre optic \acrshort{das} instead of strain gauges on a dry Bentheimer specimen. \citet{chapman2021} observed significant $E$ and $K$ but insignificant $G$ dispersion and attenuation caused by fluid pressure diffusion (FPD) in partially saturated Berea specimens featuring \acrshort{co2}-exsolution by depressurisation. \citet{tisato2021} measured dry and partially saturated Berea specimens as function of confining pressure in which the former are frequency-independent and the latter are frequency-dependent at pressures below $14$ MPa. \citet{tisato2021} also used \acrshort{kkr}-consistent models to determine the mechanism \acrshort{wiff} but was unable to distinguish between \say{squirt-flow} and patchy-saturation as sub-mechanisms. \citet{wei2021} assumed isotropy to converted $K$ and $Q_\mathrm{K}^{-1}$ from measured $E$ and $Q_\mathrm{E}^{-1}$ plus $\nu$ and $Q_\upnu^{-1}$ for two sandstones at different saturation states. Bentheimer and Bandera are opposites: $K$ dispersion is entirely absent in the former but present and increasing with saturation in the latter. Peculiar is however that only converted and no measured parameters are disclosed. \citet{han2021} measured significant \citeauthor{young1807}'s modulus $E$ and \citeauthor{poisson1827}'s ratio $\nu$ transformed to P- and S-wave velocities $V_\mathrm{P}$ and $V_\mathrm{S}$ dispersion at seismic and insignificant dispersion from seismic to ultrasonic frequencies for oil- and glycerine-saturated tight sandstones. \citet{he2021} extended on previous studies by introducing a squirt-flow extended patchy saturation model: $V_\mathrm{P}$ is both measured and modelled but $Q_\mathrm{V_{P}}^{-1}$ is only modelled. \citet{ma2021} studied $V_\mathrm{P}$ of water- and glycerine-saturated specimens. \citet{yin2021} appears to be a reiteration of \citet{yin2019}. \citet{riabokon2021a} studied the non-linearity of $286$ dry specimens at different strain amplitudes where \acrfull{ecp} and \acrfull{lv} measured axial and radial strains. \citet{lu2021phd} measured axial and shear strains of artificial specimens (3D printed and otherwise) at different stress and loading conditions from $0.01$ to $20$ Hz using a pair of \acrfull{lds}. \citet{chapman2022,chapman2022b} questioned the assumed adiabatic (thermodynamically unrelaxed) and instead argued for isothermal (thermodynamically relaxed) interaction between multiple fluid phases by studying microscopic gas bubbles. Evident $E$, $K$, and $G$ dispersion and $Q^{-1}_\mathrm{E}$, $Q^{-1}_\mathrm{K}$, and $Q^{-1}_\mathrm{G}$ attenuation peaks at $\sim100$ Hz are interpreted as thermodynamic relaxed-unrelaxed transitions. $G$ and $Q^{-1}_\mathrm{G}$ are possibly affected by pore-scale heterogeneities as experiments and numerics coincide. \citet{yurikov2022} is a reiteration of \citet{yurikov2021a,yurikov2021b}. \citet{he2022} combined \acrshort{fo} and \acrshort{pt} to calculate and measure $V_\mathrm{P}$ for three dry, brine-, and glycerine-saturated sandstones as a function of effective pressure. Pore microstructure significance is implied by simultaneously increasing effective pressure and decreasing dispersion as well as simulated by a triple porosity model with combined \say{squirt-flow} effects. Triple signifies equant, intermediate, and compliant pores with inhomogeneous aspect ratio distributions. \citet{mews2022} studied the impact of strain amplitude on $E$ for a water-saturated Bentheimer specimen as a function of effective stress. Increasing order of harmonics were also briefly analyzed in an attempt to understand non-linearities and non-elastic mechanisms coupled to strain amplitude that are fundamental for static-dynamic property bridging \citep{fjaer2019}. \citet{han2022} described measured pressure and frequency dependencies of an oil-saturated sandstone with a modified \say{squirt-flow} model assuming a triple porosity model similar to \citet{he2022}. 

    \item[\textbf{Carbonates}]\citet{spencer1981} measured $E$ and $Q_\mathrm{E}^{-1}$ of Spergen specimens at different saturation and temperature conditions. \citeauthor{spencer1981}'s vacuum-dried specimen is the stiffest and least dispersive and attenuative but his water-saturated specimens are increasingly attenuative and dispersive yet softer from $2\degree$C to $25\degree$C. \citet{paffenholzetal1989} also observed $E$ and $G$ stiffening with corresponding $Q_\mathrm{E}^{-1}$ and $Q_\mathrm{G}^{-1}$ weakening as a function of decreasing saturation. \citet{batzle2006} studied P- and S-wave velocities as a function of frequency and temperature for heavy oil-saturated Uvalde specimens that decreased in velocities with increasing temperature. \citet{adam2006} explored \citeauthor{gassmann1951}'s theory applicability to $G$ and $K$ dispersion in what was claimed the first controlled laboratory experiments on carbonates at seismic frequencies. \citet{behura2007} saw significant $Q_\mathrm{G}^{-1}$ temperature dependence (and thus viscosity dependence) in bitumen-saturated Uvalde specimens similar to the observations of \citet{spencer2013} for McMurray bitumen sand. \citet{adam2009} calculated $Q_\mathrm{K}^{-1}$, $Q_\mathrm{P}^{-1}$, and $Q_\mathrm{G}^{-1}$ from measured $Q_\mathrm{E}^{-1}$ and the complex $\nu$. In contrast to sandstone observations, $Q_\mathrm{E}^{-1}$ is higher when dry rather than fully water-saturated: $Q_\mathrm{E}^{-1}$ is frequency-dependent in both dry and saturated scenarios. \citet{zhao2014,zhao2015} are the same study presented twice: \acrshort{fo}, \acrshort{dars}, and \acrshort{pt} are combined to measure $V_\mathrm{P}$ and $V_\mathrm{S}$ in partially saturated specimens across seismic, sonic, and ultrasonic frequencies. \citet{mikhaltsevitch2016a} also studied \citeauthor{gassmann1951}'s theory applicability for elastic moduli prediction and the influence of partial water saturation on elastic and anelastic properties of Savoinni\`ere specimens. \citet{borgomano2017} found dispersion at around $200$ Hz affecting all moduli but $G$ in water-saturated Lavoux specimens. \citet{mikhaltsevitch2019a,mikhaltsevitch2019b,mikhaltsevitch2022} demonstrated that non-ideal boundary conditions (enforced by varying dead volumes) caused significant dispersion in fully decane-saturated Savoinni\`ere specimens if the dead volume exceeds the pore space. \citet{borgomano2019} interpreted the observed dispersion and attenuation of Coquina, Rustrel, and Indiana (intact and thermally cracked) specimens in terms of transitions between drained, undrained and unrelaxed fluid-flow regimes at water and glycerine-saturated conditions. \citeauthor{biot1956b}-\citeauthor{gassmann1951} theory consistency is proven at seismic frequencies. Pore type is correlated to \say{squirt-flow} dispersion absent in rocks featuring intragranular microporosity and present in rocks featuring cracked intergranular cement and uncemented grain contacts at seismic and sonic frequencies. \citet{li2020a} also studied a tight carbonate at various degrees of saturation devoid of any noteworthy attenuation and dispersion features beyond an accretion of $Q_\mathrm{E}^{-1}$ towards $1000$ Hz. \citet{tan2020} complemented \citet{mikhaltsevitch2019b} with a $1$D poroelastic model based on \citet{muller2016} that quantified the dead volume dependence predicted by \citet{pimienta2016,sun2019}. \citet{ikeda2021} measured increasing $E$ and decreasing $Q_\mathrm{E}^{-1}$ with increasing water-saturation for a polymineralic carbonate. \citet{riabokon2021b} also studied the non-linearity of dry carbonates. \citet{mikhaltsevitch2022} is the continuation of \citet{mikhaltsevitch2019a,mikhaltsevitch2019b} whose results are also modelled by modified \citeauthor{gassmann1951} theory. \citet{sun2022} studied the impact of partial saturation by (i) drying and (ii) imbibition: P-wave dispersion (and attenuation) is significant for (i) but not for (ii) whereas S-wave dispersion (and attenuation) is insignificant for both (i) and (ii) at $>80\%$ \acrshort{rh}. Mesoscopic \acrshort{wiff} controlled by geometry and pore fluid distribution is the main mechanism causing P-wave dispersion. \citet{gallagher2022} hydrostatically investigated unfractured and fractured Rustrel specimens \citep{borgomano2019} at different effective pressures in triaxial and undrained conditions to better understand the effect of fractures and validate computational fracture models. Aside from a local negative phase shift for the fractured specimen at saturated conditions, no attenuation is observed at dry conditions.

    \item[\textbf{Shales}] \citet{batzle2005,batzle2014} considered clay particle interactions with bound water responsible for shale attenuation and dispersion. \citet{duranti2005} discovered that dispersion in shales occurs between sonic and ultrasonic frequencies in addition to the seismic band being characterized by nearly constant attenuation. \citet{hofmann2006,sarker2010} studied the frequency dependence of water and glycerine saturated Mancos specimens. Despite proven able to measure $Q_\mathrm{E}^{-1}$ \citep{batzle2006}, \citet{hofmann2006,sarker2010} solely focused on $E$ and $\nu$ converted into $V_\mathrm{P_{0}}$, $V_\mathrm{P_{45}}$, $V_\mathrm{P_{90}}$, $V_\mathrm{S_{0}}$, and $V_\mathrm{S_{90}}$ via $C_{ij}$. \citet{delle_piane_2011,delle_piane_2014} made a two-fold discovery: (i) attenuation is greater normal to bedding, and (ii) partial saturation increases the attenuation (likely due to micro and mesoscopic flow). Failure to disclose $E$ and $\nu$ was due to deficient calibration of the absolute displacement signals related to different amplification factors. \citet{mikhaltsevitch2015} linked seismic and ultrasonic measurements of dry and wet $0\degree$ Eagle Ford specimens found to be non-dispersive and non-attenuative by assuming isotropy. Both measurement types were concluded to be in the high-frequency regime due to partial saturation. \citet{huang2015} measured Mancos dispersion and attenuation: $E_\mathrm{V}>E_\mathrm{H}$ and $Q_\mathrm{E_V}^{-1}>Q_\mathrm{E_H}^{-1}$. Low $Q_E$ was also measured for Xianjing specimens. \citet{mikhaltsevitch2017} studied dispersion and attenuation of fully saturated Wellington ($0$, $45$, and $90\degree$) and a partially saturated Mancos ($0$ and $90\degree$) specimens. $Q_\mathrm{E_{45}}^{-1}>Q_\mathrm{E_{H}}^{-1}>Q_\mathrm{E_{V}}^{-1}$ appears to be a clandestine trend within the scattered Wellington data $Q_\mathrm{E_{x}}^{-1}<0.0075$ in which $x$ refers to subscripts $V$, $H$, and $45$. Trend or no trend, \citet{mikhaltsevitch2017} measured Wellington at non-dispersive frequencies due to the low $Q_\mathrm{E_{x}}^{-1}$ devoid of any noteworthy peaks. Like Wellington $Q_\mathrm{E_{V}}^{-1}$ and $Q_\mathrm{E_{H}}^{-1}$, Mancos $Q_\mathrm{E_{V}}^{-1}$, and $Q_\mathrm{E_{H}}^{-1}$ are also insignificant and unnoteworthy at \acrshort{rh}s below $97.5\%$. Unlike at other \acrshort{rh}s, $97.5\%$ features $Q_\mathrm{E_{V}}^{-1}$ and $Q_\mathrm{E_{H}}^{-1}$ peaks. $Q_\mathrm{E_{V}}^{-1}>Q_\mathrm{E_{H}}^{-1}$ and $E_\mathrm{H} < E_\mathrm{V}$ universally applies to \citet{mikhaltsevitch2017} who also observed gradual softening accompanied by monotonically decreasing $E_\mathrm{V}$ and $E_\mathrm{H}$ with increasing saturation. \citet{mikhaltsevitch2016w,mikhaltsevitch2018} are based on the Wellington experiments summarized in \citet{mikhaltsevitch2017}. \citet{mikhaltsevitch2021} is the recent extension of the Mancos experiments \citep{mikhaltsevitch2016m,mikhaltsevitch2017m,mikhaltsevitch2017}. Anisotropic permeability is their explanation to peak $Q_\mathrm{E_{V}}^{-1}$ and $Q_\mathrm{E_{H}}^{-1}$ occurring at different frequencies. Global flow ensured by the drained-undrained transition explains the presence or absence of dispersion at the probed frequencies. Wellington is also studied by \citet{mikhaltsevitch2021b} but at different saturation states with similar dispersion features (and explanation) as Mancos \citep{mikhaltsevitch2021}: peak $Q_\mathrm{E_{V}}^{-1}$ and $Q_\mathrm{E_{H}}^{-1}$ at $2$ and $6$ Hz. \citet{mikhaltsevitch2021b} also includes the Eagle Ford experiments by \citet{chavez2019}. \citet{xiao2021} compared different attenuation mechanisms based on characteristic frequencies for $0$ and $90\degree$ specimens of an unnamed shale. \citet{rorheim2022phd} observed non-simultaneous uniaxial-stress $E_\mathrm{V}$ and uniaxial-strain $C_{33}$ behaviours as well as decreasing $Q_\mathrm{E_V}^{-1}$ and $Q_\mathrm{C_{33}}^{-1}$ with decreasing saturation for five differently oriented and saturated Pierre specimens. $E_\mathrm{V}$ is doubled once and $\nu_\mathrm{VH}$ is halved twice with decreasing saturation. $E_\mathrm{V}<E_{45}<E_\mathrm{H}$ is continuous while initially $Q_\mathrm{E_{H}}^{-1}<Q_\mathrm{E_{45}}^{-1}<Q_\mathrm{E_{V}}^{-1}$ ultimately becomes $Q_\mathrm{E_{V}}^{-1}<Q_\mathrm{E_{H}}^{-1}<Q_\mathrm{E_{45}}^{-1}$ with increasing frequency. \acrshort{fo}-measured $C_{33}$ and \acrshort{pt}-measured $V_\mathrm{P_0}$ decrease simultaneously with decreasing saturation despite the frequency gap. No $Q_\mathrm{E_V}^{-1}$ attenuation peaks but $E_\mathrm{V}$ dispersion were noticed. \citet{li2022} examined the impact of maturity on attenuation of four different oven-dried $0\degree$ specimens: Barnett, Eagle Ford, Mancos, and Jungar. Constant dispersion and insignificant attenuation are features of the three former however the latter feature strong dispersion and significant dispersion which decreases with increasing pressure. The controlling factor is not thermal maturity or clay content but instead the geochemical index attributed to viscous friction between inorganic grains and organic matter. \citet{long2022} excluded capillary effects and \acrshort{wiff} but included viscoelastic bulk and shear moduli as plausible bound water attenuation mechanisms while researching seven unnamed shale specimens.
    
    Although unable to record $Q_\mathrm{E}^{-1}$ measurements, \citet{szewczyk2016} observed strong softening at seismic and hardening counteracting this softening at ultrasonic frequencies (due to increasing dispersion) with increasing water saturation in Mancos specimens. $\nu$ also strongly increased but appeared nearly frequency-independent. Types of shales also differ in stress sensitivity during hydrostatic loading which also differs at seismic and ultrasonic frequencies. \citet{li2017} referred to \citet{huang2015} when measuring $E$ and $\nu$ dispersion of an unnamed field shale. \citet{szewczyk2018a} assumed variable and linearly decreasing $C_{ij}^\mathrm{dry}$ with increasing \acrshort{rh} for anisotropic \citeauthor{gassmann1951}'s theory to be applicable to Mancos. \citet{lozovyi2018phd} succeeded \citet{szewczyk2017phd} by comparing the static and dynamic stiffness of Opalinus Clay and three other anonymous specimens \citep{lozovyi2018arama,lozovyi2018,lozovyi2019b}. \citet{chavez2019} conducted \acrshort{fo} creep tests on three dry sedimentary rocks at constant $2$ Hz for $120$ hours. Creep effects significantly impacted the moduli for all three specimens but the shale was most affected. \citet{mikhaltsevitch2021} questioned the applicability of anisotropic \citeauthor{gassmann1951} theory to Mancos proposed by \citet{szewczyk2018a} on the basis that $C_{ij}^\mathrm{sat}>C_{ij}^\mathrm{dry}$ is repudiated by the non-linearity of the decreasing $C_{44}$ and $C_{66}$ with saturation. Be the fact that $C_{44}^\mathrm{sat}<C_{44}^\mathrm{dry}$ and $C_{66}^\mathrm{sat}<C_{66}^\mathrm{dry}$ as it may, the non-linearity is questioned (only applies to $C_{44}$ and $C_{66}$ at $9\%$ \acrshort{rh}). \citet{rorheim2021c} performed \acrshort{fo} measurements as functions of frequency as well as time on a \acrshort{co2}-exposed Draupne specimen where $E_\mathrm{V}$ and $\nu_\mathrm{VH}$ insignificantly changed over $575$ hours. Neither \acrshort{pt} measured $V_\mathrm{P_0}$ nor $V_\mathrm{S_0}$ significantly changed perhaps due to Draupne's low calcite content. \citet{rorheim2021a} combined \acrshort{fo} and \acrshort{pt} at elevated temperatures for $0$ and $90\degree$ Pierre specimens: \acrshort{fo} measured $E_\mathrm{V}$ and $E_\mathrm{H}$ dispersion shifted towards higher frequencies while \acrshort{fo} converted and \acrshort{pt} measured $V_\mathrm{P_0}$, $V_\mathrm{P_{90}}$, $V_\mathrm{S_0}$, and $V_\mathrm{S_{90}}$ oppositely decreased and increased with temperature. Two simple frequency-dependent bound water models were also proposed.
\end{enumerate}

\nomenclature{$V_\mathrm{P}$}{P-wave velocity.}
\nomenclature{$V_\mathrm{S}$}{S-wave velocity.}
\nomenclature{$Q_\upnu^{-1}$}{Inverse quality factor of \citeauthor{poisson1827}'s ratio $\nu$.}
\nomenclature{$V_\mathrm{P_0}$}{P-wave velocity normal to bedding.}
\nomenclature{$V_\mathrm{P_{45}}$}{P-wave velocity oblique ($45\degree$) to bedding.}
\nomenclature{$V_\mathrm{P_{90}}$}{P-wave velocity perpendicular to bedding.}
\nomenclature{$V_\mathrm{S_0}$}{S-wave velocity normal to bedding.}
\nomenclature{$V_\mathrm{S_{90}}$}{S-wave velocity perpendicular to bedding.}
\nomenclature{$Q_\mathrm{E_V}^{-1}$}{Inverse quality factor of \citeauthor{young1807}'s modulus $E_\mathrm{V}$ assuming \acrshort{ti} symmetry.}
\nomenclature{$Q_\mathrm{E_{45}}^{-1}$}{Inverse quality factor of \citeauthor{young1807}'s modulus $E_{45}$ assuming \acrshort{ti} symmetry.}
\nomenclature{$Q_\mathrm{E_H}^{-1}$}{Inverse quality factor of \citeauthor{young1807}'s modulus $E_\mathrm{H}$ assuming \acrshort{ti} symmetry.}

Numerical studies supplementing experimental ones are becoming customary. Experimentally proven by \citet{tisato2012}, \citet{quintal2011a} and \citet{quintal2011b} were the numerical impetus of \citet{tisato2013,tisato2014,kuteynikova2014,quintal2017,chapman2018,hunziker2018,quintal2019,alkhimenkov2020,lissa2020,lissa2020seg,lissa2021egu,lissa2021,chapman2021,alkhimenkov2021,chapman2022,gallagher2022,alkhimenkov2022} in terms of attenuation modelling. \citet{zhang2012,das2019,janicke2019} used similar \acrfull{drp} approaches to study attenuation mechanisms and fluid-solid interactions. \citet{balcewicz2021} used \acrshort{drp} to study other properties. Focusing on enhancing an existing \acrshort{fo} approach, \citet{sun2018} tried to alleviate the experimental destructive subsonic resonances via numerical modelling to redefine their apparatus design. \citet{sun2018} achieved this by inverting \citeauthor{tittmann1977}'s principle manifested by \citeauthor{nakagawa2011}'s loaded resonator: the shorter the apparatus, the higher the characteristic frequency. \citet{sun2018} inspired \citet{liu2019} to investigate the cause and effect of subsonic resonances for the apparatus described by \citet{mikhaltsevitch2014} in a similar manner. \citet{sun2019} $3$D modelled the drained-undrained transition for the frequency-dependent elastic moduli and attenuation. In order to inaugurate an upper frequency-limit, \citet{borgomano2020,li2020b} adopted the \acrshort{comsol}-based numerical approach by \citet{sun2018} to deduce at which frequencies their apparatus resonates. Unlike \citet{sun2018,borgomano2020}, \citet{li2020b} included jointing conditions in their model because the stress-field is non-homogeneous. As a result, elastic properties depend on strain gauge position. Be aware however that numerical need not equal experimental due to the ideal versus non-ideal states predicament. \citet{ikeda2021} compared experimental and numerical results: $10\%$ separates \acrshort{drp}-derived and \acrshort{fo}-measured $E$.

\begin{figure}[H]
    \centering
    \includegraphics[width=\linewidth]{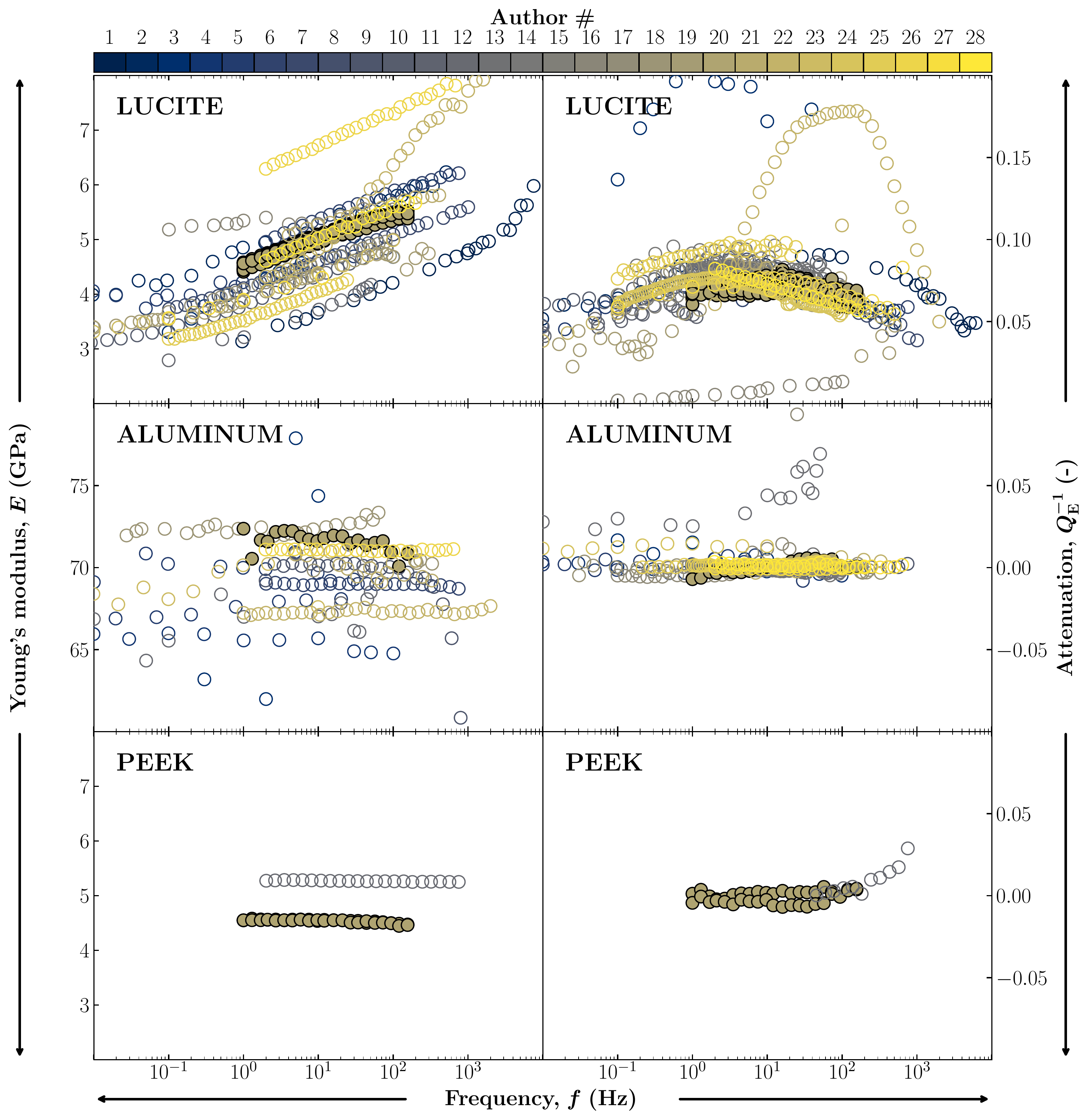}
    \caption{\citeauthor{young1807}'s modulus $E$ and attenuation $Q_\mathrm{E}^{-1}$ versus frequency $f$. Color-bars denote authors according to Author \# listed in Table~\ref{tab:inv_q_lit}. Enhanced distinguishability is ensured by solid symbols for the bolded \textbf{\citet{rorheim2019}} and open symbols for all other authors.}
    \label{fig:inv_q_lit}
\end{figure}

\acrshort{fo} apparatuses are proven by their ability to measure attenuation of standard materials often used for calibration purposes: lucite, aluminium, and \acrshort{peek}. Other calibration materials exist (e.g. \citet{borgomano2020} calibrated for glass and gypsum in addition to lucite, while \citet{brunner2003} used a unspecified \say{viscoelastic structure}) but are excluded due to their rarity. To the best of their knowledge, \citet{rorheim2019} compiled and studied all published \acrshort{fo} measurements involving these three materials (Table~\ref{tab:inv_q_lit}) to validate their findings by literature comparison (Figure~\ref{fig:inv_q_lit}\footnote{Please be advised that all external measurements from other authors are digitized with variable resolution during which resonance-affected data are omitted. Measurements at non-ambient temperatures \citep{iwasaki1980,iwasaki1984,esnouf1981,wei2002,riviere2003} are also omitted.}). Lucite is dispersive opposed to non-dispersive aluminium and \acrshort{peek}. Questioned is the calibration applicability of high modulus aluminium dissimilar to rocks but not that of low moduli lucite and \acrshort{peek} similar to rocks. \citet{batzle2006} recognized that the composition of lucite can be variable from batch to batch. In fact, \citet{bonner2019} claimed that the hardener-resin ratio determines the quality of the material. Unlike the inexpensive resin, the hardener is expensive. It is thus common for resin to be the dominant component of the hardener-resin ratio. This primarily affects the absolute mechanical properties but not the overall frequency dispersion characteristics \citep{saltiel2017}. Figure~\ref{fig:inv_q_lit} possibly features this effect as \citet{sun2018} is an evident outlier compared to all others.

\begin{table}[H]
    \caption{All known $Q_\mathrm{E}$ measurements of lucite, aluminium, and \acrshort{peek} alphabetically sorted and categorized by author(s) and specimen material, respectively. Color-bars based on Author \# and the (\texttt{{\color{civ_col1}c}{\color{civ_col2}i}{\color{civ_col3}v}{\color{civ_col4}i}{\color{civ_col5}d}{\color{civ_col6}i}{\color{civ_col7}s}}) color-map are used instead of legends. \textbf{\citet{rorheim2019}} is in bold because this is based on that study in which all known \acrshort{fo} studies were compared.}
    \centering
    \begin{tabular}{lcccc}
        \toprule[1.5pt]
        \multirow{2}{*}{Author(s)}                      & \multirow{2}{*}{Author \#}    & \multicolumn{3}{c}{Specimen(s)}                                                                           \\
                                                                              \cmidrule(lr){3-5}
                                                        &                               & Lucite                        & Aluminium                             & \acrshort{peek}                   \\
        \midrule
        \citet{batzle2006}                              & {\color{col1}$1$}             & {\color{green}\cmark}         & {\color{red}\xmark}                   & {\color{red}\xmark}               \\
        \citet{borgomano2020}                           & {\color{col2}$2$}             & {\color{green}\cmark}         & {\color{red}\xmark}                   & {\color{red}\xmark}               \\
        \citet{cao2021}                                 & {\color{col3}$3$}             & {\color{green}\cmark}         & {\color{green}\cmark}                 & {\color{red}\xmark}               \\
        \citet{cherry1996}                              & {\color{col4}$4$}             & {\color{red}\xmark}           & {\color{green}\cmark}                 & {\color{red}\xmark}               \\
        \citet{fliedner2021}                            & {\color{col5}$5$}             & {\color{green}\cmark}         & {\color{green}\cmark}                 & {\color{red}\xmark}               \\
        \citet{huang2015}                               & {\color{col6}$6$}             & {\color{green}\cmark}         & {\color{red}\xmark}                   & {\color{green}\cmark}             \\
        \citet{katzmann2019}                            & {\color{col7}$7$}             & {\color{green}\cmark}         & {\color{red}\xmark}                   & {\color{red}\xmark}               \\
        \citet{koppelmann1958} via \citet{lakes2009}    & {\color{col8}$8$}             & {\color{green}\cmark}         & {\color{red}\xmark}                   & {\color{red}\xmark}               \\
        \citet{li2017}                                  & {\color{col9}$9$}             & {\color{red}\xmark}           & {\color{green}\cmark}                 & {\color{red}\xmark}               \\
        \citet{lienertetal1990}                         & {\color{col10}$10$}           & {\color{green}\cmark}         & {\color{red}\xmark}                   & {\color{red}\xmark}               \\
        \citet{liu1983}                                 & {\color{col11}$11$}           & {\color{green}\cmark}         & {\color{red}\xmark}                   & {\color{red}\xmark}               \\
        \citet{lu2021phd}                               & {\color{col12}$12$}           & {\color{green}\cmark}         & {\color{green}\cmark}                 & {\color{red}\xmark}               \\
        \citet{madonna2011}                             & {\color{col13}$13$}           & {\color{green}\cmark}         & {\color{green}\cmark}                 & {\color{red}\xmark}               \\
        \citet{madonna2013}                             & {\color{col14}$14$}           & {\color{green}\cmark}         & {\color{green}\cmark}                 & {\color{red}\xmark}               \\
        \citet{mccann2019msc}                           & {\color{col15}$15$}           & {\color{green}\cmark}         & {\color{green}\cmark}                 & {\color{red}\xmark}               \\
        \citet{mikhaltsevitch2016a}                     & {\color{col16}$16$}           & {\color{green}\cmark}         & {\color{red}\xmark}                   & {\color{red}\xmark}               \\
        \citet{nakagawa2013LF}                          & {\color{col17}$17$}           & {\color{green}\cmark}         & {\color{red}\xmark}                   & {\color{red}\xmark}               \\
        \citet{paffenholzetal1989}                      & {\color{col18}$18$}           & {\color{red}\xmark}           & {\color{green}\cmark}                 & {\color{red}\xmark}               \\
        \citet{pimienta2015b}                           & {\color{col19}$19$}           & {\color{green}\cmark}         & {\color{red}\xmark}                   & {\color{red}\xmark}               \\
        \textbf{\citet{rorheim2019}}                    & {\color{col20}$20$}           & {\color{green}\cmark}         & {\color{green}\cmark}                 & {\color{green}\cmark}             \\
        \citet{spencer1981}                             & {\color{col21}$21$}           & {\color{green}\cmark}         & {\color{green}\cmark}                 & {\color{red}\xmark}               \\
        \citet{sun2018}                                 & {\color{col22}$22$}           & {\color{green}\cmark}         & {\color{red}\xmark}                   & {\color{red}\xmark}               \\
        \citet{takei2011}                               & {\color{col23}$23$}           & {\color{green}\cmark}         & {\color{green}\cmark}                 & {\color{red}\xmark}               \\
        \citet{tisato2012}                              & {\color{col24}$24$}           & {\color{green}\cmark}         & {\color{green}\cmark}                 & {\color{red}\xmark}               \\
        \citet{tisato2016}                              & {\color{col25}$25$}           & {\color{green}\cmark}         & {\color{red}\xmark}                   & {\color{red}\xmark}               \\
        \citet{yao2013phd}                              & {\color{col26}$26$}           & {\color{green}\cmark}         & {\color{green}\cmark}                 & {\color{red}\xmark}               \\
        \citet{yee1982}                                 & {\color{col27}$27$}           & {\color{green}\cmark}         & {\color{red}\xmark}                   & {\color{red}\xmark}               \\
        \citet{yin2017}                                 & {\color{col28}$28$}           & {\color{green}\cmark}         & {\color{green}\cmark}                 & {\color{red}\xmark}               \\
        \bottomrule[1.5pt]
    \end{tabular}%
    \label{tab:inv_q_lit}
\end{table}

\begin{sidewaystable}
\centering
\footnotesize
\captionsetup{size=footnotesize}

\caption{Sonic frequency probing techniques with primary focus on \acrfull{rb} but secondary also on \acrfull{ptu}, \acrfull{ghrc}, \acrfull{rus}, and \acrfull{dars}.}

\label{tab:review}

\begin{threeparttable}
    \begin{tabular}{llllll}
    \toprule[1.5pt]
    Author(s)                           & Technique(s)      & Frequency (Hz)                            & Length (cm)               & Diameter (cm)                 & Parameter(s) \\
    \midrule
    \citet{batzle2006}                  & \acrshort{fo}     & $5\phantom{.0}$E$0-2.5\phantom{0}$E$3$    &                           &                               & {$E$}, {$V_\mathrm{P}$}, {$V_\mathrm{S}$} {$Q_\mathrm{E}$} \\
    \citet{birch1938}                   & \acrshort{rb}     & $1.4$E$2-4.5\phantom{0}$E$3$              & {$244\phantom{.}$}        & {$23$}                        & {$Q$} \\
    \citet{born1941}                    & \acrshort{rb}     & $9.3$E$2-1.28$E$4$                        & {$14\phantom{.0}-124$}    & {$12$}                        & {${\delta}={\pi}Q^{-1}$} \\
    \citet{bourbie1985}                 & \acrshort{rb}     & $3\phantom{.0}$E$3-5\phantom{.00}$E$3$    &                           &                               & {$V_\mathrm{E}$}, {$Q$} \\
    \citet{cadoret1995}                 & \acrshort{rb}     & $1\phantom{.0}$E$3$                       & {$110\phantom{.}$}        & {$8$}                         & {$V_\mathrm{E}$}, {$V_\mathrm{S}$}, {$V_\mathrm{P}$} \\
    \citet{cadoret1998}                 & \acrshort{rb}     & $1\phantom{.0}$E$3$                       & {$110$}                   & {$8$}                         & {$Q_\mathrm{G}$}, {$Q_\mathrm{E}$} \\
    \citet{gardner1964}                 & \acrshort{rb}     & $2\phantom{.0}$E$3-3\phantom{.00}$E$3$    & {$5\phantom{.00}-30.0$}   & {$5$}                         & {$\delta_\mathrm{E}$}, {$\delta_\mathrm{S}$} \\
    \citet{goldberg1989}                & \acrshort{rb}     & $5\phantom{.0}$E$3-25\phantom{.0}$E$3$    & {$25$}                    & {$2.5$}                       & {$Q_\mathrm{P}$} \\
    \citet{ide1935}                     & \acrshort{rb}     & $4\phantom{.0}$E$3-12\phantom{.0}$E$3$    & {$25$}                    & {$5.1$}                       & {$E$} \\
    \citet{harris2005}                  & \acrshort{dars}   & $1\phantom{.0}$E$3-2\phantom{.00}$E$3$    &                           &                               & $K$, $Q_\mathrm{K}$    \\
    \citet{jones1983}                   & \acrshort{rb}     & $1.7$E$2-3.4\phantom{0}$E$3$              &                           &                               & {$V_\mathrm{S}$}, {$Q_\mathrm{G}$} \\
    \citet{lucet1991}                   & \acrshort{rb}     & $5\phantom{.0}$E$3-2\phantom{.00}$E$4$    & {$25\phantom{.0}-30$}     & {$2.5$}                       & {$V_\mathrm{E}$}, {$V_\mathrm{P}$}, {$Q_\mathrm{E}$}, {$Q_\mathrm{G}$} \\
    \citet{lucet1992}                   & \acrshort{rb}     & $3\phantom{.0}$E$3-1\phantom{.00}$E$4$    & {$30$}                    & {$2.5$}                       & {$Q_\mathrm{E}$} \\
    \citet{lucet2006}                   & \acrshort{rb}     & $2\phantom{.0}$E$3-2\phantom{.00}$E$4$    &                           &                               & {$Q_\mathrm{E}$}, {$V_\mathrm{S}/V_\mathrm{US}$}, {$V_\mathrm{PP}/V_\mathrm{PFB}$} \\
    \citet{mccann2014}                  & \acrshort{ptu}    & $1\phantom{.0}$E$3-1\phantom{.00}$E$4$    & {$60$}                    & {$6.9$}                       & {$V_\mathrm{P}$}, {$Q_\mathrm{P}$}, {$T$} \\
    \citet{murphy1982}                  & \acrshort{rb}     & $3\phantom{.0}$E$2-1.4\phantom{0}$E$4$    & {$20\phantom{.0}-100$}    &                               & {$V_\mathrm{S}$}, {$V_\mathrm{E}$}, {$Q_\mathrm{E}$}, {$Q_\mathrm{G}$} \\
    \citet{murphy1984}                  & \acrshort{rb}     & $5\phantom{.0}$E$3$                       & {$20\phantom{.0}-25$}     & {$19$}                        & {$V_\mathrm{S}$}, {$V_\mathrm{E}$}, {$Q_\mathrm{E}$}, {$Q_\mathrm{G}$} \\
    \citet{nakagawa2010}                & \acrshort{shrb}   & $3.5$E$2-2.35$E$3$                        & {$6.2\phantom{0}$}        & {$3.75$}                      & {$E$}, {$G$}, {$\nu$}, {$V_\mathrm{P}$}, {$V_\mathrm{S}$}, {$Q$} \\
    \citet{nakagawa2011}                & \acrshort{shrb}   & $4\phantom{.0}$E$2-2.3\phantom{0}$E$3$    & {$6.22$}                  & {$3.81$}                      & {$E$}, {$G$}, {$\nu$}, {$V_\mathrm{P}$}, {$V_\mathrm{S}$}, {$Q$} \\
    \citet{nakagawa2011a}               & \acrshort{shrb}   & $3\phantom{.0}$E$2-1.5\phantom{0}$E$3$    & {$7.62$}                  & {$3.75$}                      & {$E$}, {$G$}, {$\nu$}, {$V_\mathrm{P}$}, {$V_\mathrm{S}$}, {$Q$} \\
    \citet{ohara_1985}                  & \acrshort{rb}     & $3\phantom{.0}$E$2-3\phantom{.00}$E$3$    & {$38$}                    & {$2.22$}                      & {$V_\mathrm{E}$}, {$V_\mathrm{S}$}, {$\delta$} \\
    \citet{priest2006}                  & \acrshort{ghrc}   & $4\phantom{.0}$E$2$                       & {$14$}                    & {$7$}                         & {$Q_\mathrm{E}$}, {$Q_\mathrm{G}$} \\
    \citet{tittmann1977}                & \acrshort{rb}     & $2.2$E$4-2.3\phantom{0}$E$4$              &                           &                               & {$Q_\mathrm{E}$}, {$Q_\mathrm{G}$} \\
    \citet{tittmann1981}                & \acrshort{rb}     & $7\phantom{.0}$E$3-9\phantom{.00}$E$3$    & {$12$}                    & {$1.5$}                       & {$Q_\mathrm{E}$} \\
    \citet{waite2011}                   & \acrshort{shrb}   & $3.6$E$2-1.6\phantom{0}$E$3$              & {$7.62$}                  & {$3.81$}                      & {$V_\mathrm{P}$}, {$V_\mathrm{S}$} \\
    \citet{wegel1935}                   & \acrshort{rb}     & $1\phantom{.0}$E$2-1\phantom{.00}$E$5$    & {$30$}                    & {$1$}                         & {$Q_\mathrm{E}$}, {$Q_\mathrm{G}$} \\ 
    \citet{winkler1979}                 & \acrshort{rb}     & $5\phantom{.0}$E$2-1.7\phantom{0}$E$3$    & {$100$}                   &                               & {$V_\mathrm{E}$}, {$V_\mathrm{S}$}, {$V_\mathrm{P}$}, {$Q_\mathrm{E}$}, {$Q_\mathrm{G}$}, {$Q_\mathrm{K}$}, {$Q_\mathrm{P}$} \\
    \citet{winkler1982}                 & \acrshort{rb}     & $5\phantom{.0}$E$2-9\phantom{.00}$E$3$    & {$100$}                   &                               & {$V_\mathrm{E}$}, {$V_\mathrm{S}$}, {$V_\mathrm{P}$}, {$Q_\mathrm{E}$}, {$Q_\mathrm{G}$}, {$Q_\mathrm{K}$}, {$Q_\mathrm{P}$} \\
    \citet{wyllie1962}                  & \acrshort{rb}     & $2\phantom{.0}$E$4$                       &                           & {$1.9\phantom{0}-2.50$}       & {$V_\mathrm{E}$}, {$V_\mathrm{S}$}, {$\nu$}, {$\delta_\mathrm{E}$}, {$\delta_\mathrm{S}$} \\
    \citet{yin1992}                     & \acrshort{rb}     & $1.6$E$2-1.8\phantom{0}$E$3$              & {$39\phantom{.0}-53$}     & {$5$}                         & {$E$}, {$Q_\mathrm{E}$} \\
    \citet{zadler2004}                  & \acrshort{rus}    & $1.4$E$2-8.8\phantom{0}$E$4$              & {$7.1$}                   & {$2.5$}                       & {$Q_\mathrm{E}$}, {$Q_\mathrm{G}$}, {$V_\mathrm{P}$}, {$V_\mathrm{S}$} \\
    \bottomrule[1.5pt]
    \end{tabular}%

\end{threeparttable}
\end{sidewaystable}

\begin{sidewaystable}

\centering
\footnotesize
\captionsetup{size=footnotesize}

\caption{Comparison of \acrshort{fo} apparatuses based on longitudinal, torsional, volumetric, and uniaxial modes. Partly based on \citet{subramaniyan2014} but improved and updated as the technique became increasingly acknowledged. Institutions are acronymised due to space limitations.}
\label{tab:FO}

\begin{threeparttable}
    \resizebox{0.9\hsize}{!}{\begin{tabular}{lllllllll}%
    \toprule[1.5pt]
    \textbf{Mode}                               & Author(s)                                     & Specimen(s)           & Force generator               & Force sensor                  & Displacement sensor                   & Frequency (Hz)                                        & Parameter(s)                                                  & Institution(s)        \\
    \midrule
    \multirow{35}{*}{\textbf{(i) Longitudinal}} & \citet{adam2009phd}                           & Lucite                & Shaker                        & Aluminium STD                 & Fibre optics                          & $5$E$0\phantom{-}-2.5\phantom{0}$E$3$                 & {$E$}, {$Q_\mathrm{E}^{-1}$}, {$\nu$}                         & \acrshort{csm}        \\
                                                & \citet{batzle2006}                            & Sandstone             & Shaker                        & Aluminium STD                 & Strain gauges\tnote{S}                & $5$E$0\phantom{-}-2.5\phantom{0}$E$3$                 & {$E$}, {$Q_\mathrm{E}^{-1}$}, {$\nu$}                         & \acrshort{csm}        \\
                                                & \citet{borgomano2020}                         & Limestone             & \acrshort{pzt} actuator       & Aluminium STD                 & Strain gauges\tnote{F}                & $4$E$-3-1\phantom{.00}$E$3$                           & {$E$}, {$Q_\mathrm{E}^{-1}$}, {$\nu$}                         & \acrshort{ens}        \\
                                                & \citet{bruckshawetal1961}                     & Various               & Coils                         & Coils                         & Coils                                 & $4$E$1\phantom{-}-1.2\phantom{0}$E$2$                 & {$E$}, {$\Delta W/W$}                                         & \acrshort{icl}        \\
                                                & \citet{cao2021}                               & Sapphire              & \acrshort{pzt} actuator       & Load cell                     & Linear encoder                        & $1$E$-1-1\phantom{.00}$E$2$                           & {$E$}, {$Q_\mathrm{E}^{-1}$}                                  & \acrshort{uo}         \\
                                                & \citet{cherry1996}                            & Sandstone             & \acrshort{pzt} actuator       & Aluminium STD                 & Interferometer                        & $1$E$-3-1\phantom{.00}$E$2$                           & {$E$}, {$Q_\mathrm{E}^{-1}$}                                  & \acrshort{ucb}        \\
                                                & \citet{delle_piane_2014}                      & Shale                 & Motor                         & Load cell                     & Strain gauge cantilever               & $1$E$-1-1\phantom{.00}$E$2$                           & {$E$}, {$Q_\mathrm{E}^{-1}$}, $\nu$                           & \acrshort{icl}        \\
                                                & \citet{david2012}                             & Sandstone             & \acrshort{pzt} actuator       & Aluminium STD                 & Strain gauges\tnote{S}                & $1$E$-3-3\phantom{.00}$E$2$                           & {$E$}, {$Q_\mathrm{E}^{-1}$}                                  & \acrshort{eth}        \\
                                                & \citet{fliedner2019}                          & Schist                & \acrshort{clsc} actuator      & Linear potentiometer          & \acrshort{lvdt}s                      & $5$E$-2-1.5\phantom{0}$E$1$                           & {$E$}, {$Q_\mathrm{E}^{-1}$}                                  & \acrshort{rice}       \\
                                                & \citet{fortin2005}                            & Sandstone             & \acrshort{pzt} actuator       & Load cell                     & Strain gauges\tnote{F}                &                                                       & {$E$}, {$\nu$}                                                & \acrshort{ens}        \\
                                                & \citet{huang2015}                             & Shale                 & \acrshort{pzt} actuator       & Aluminium STD                 & Strain gauges\tnote{S}                & $2$E0$\phantom{-}-8\phantom{.00}$E$2$                 & {$E$}, {$\nu$}, {$A$}, {$\theta$}                             & \acrshort{uh}         \\
                                                & \citet{ikeda2020}                             &                       & \acrshort{pzt} actuators      & Titanium STD                  & Capacitive sensor                     & $1$E$-1-1\phantom{.00}$E$2$                           & {$E$}, {$Q_\mathrm{E}^{-1}$}                                  & \acrshort{uta}        \\
                                                & \citet{katzmann2019}                          & Lucite                & \acrshort{pzt} actuator       & Aluminium STD                 & Strain gauges\tnote{F}                & $1$E$-2-1\phantom{.00}$E$3$                           & {$E$}, {$Q_\mathrm{E}^{-1}$}, {$\nu$}, {$Q_\upnu^{-1}$}       & \acrshort{us}         \\
                                                & \citet{lienertetal1990}                       & Sandstone             & Shaker                        & \acrshort{pzt} transducer     & Capacitive probe                      & $1$E$-1-1\phantom{.00}$E$2$                           & {$E$}, {$Q_\mathrm{E}^{-1}$}                                  & \acrshort{uoh}        \\
                                                & \citet{li2017}                                & Shale                 & \acrshort{pzt} actuator       & Aluminium STD                 & Strain gauges\tnote{S}                & $2$E$0\phantom{-}-8\phantom{.00}$E$2$                 & {$E$}, {$\nu$}, {$A$}, {$\theta$}                             & \acrshort{uh}         \\
                                                & \citet{li2019}                                & Sandstone             & Shaker                        & Aluminium STD                 & Strain gauges                         & $5$E$0\phantom{-}-2\phantom{.00}$E$3$                 & {$E$}, {$\nu$}                                                & Sinopec               \\
                                                & \citet{lu2021phd}                             & Sandstone             & MTS actuator                  & Load cell                     & Laser                                 & $1$E$-2-5\phantom{.00}$E$1$                           & {$E$}                                                         & \acrshort{uofa}       \\
                                                & \citet{mccarthy2016}                          & Polycrystalline ice   & Servomechanical-actuator      &                               & \acrshort{lvdt}s                      & $1$E$-4-1\phantom{.00}$E$-1$                          & {$E$}, {$Q_\mathrm{E}^{-1}$}                                  &                       \\
                                                & \citet{madonna2013}                           & Sandstone             & \acrshort{pzt} actuator       & Aluminium STD                 & \acrshort{lvdt}s                      & $1$E$-2-1\phantom{.00}$E$2$                           & {$E$}, {$Q_\mathrm{E}^{-1}$}                                  & \acrshort{eth}        \\
                                                & \citet{mikhaltsevitch2014}                    & Sandstone             & \acrshort{pzt} actuator       & Aluminium STD                 & Strain gauges\tnote{S}                & $1$E$-1-4\phantom{.00}$E$2$                           & {$E$}, {$Q_\mathrm{E}^{-1}$}, {$\nu$}                         & \acrshort{cu}         \\
                                                & \citet{nakagawa2013LF}                        & Sandstone             & \acrshort{pzt} actuator       & Load cell                     & Gauges attached to two rings          & $1$E$-3-1\phantom{.00}$E$2$                           & $E$, $G$, $Q_\mathrm{E}^{-1}$, $Q_G^{-1}$                     & \acrshort{lbl}        \\
                                                & \citet{paffenholzetal1989}                    & Various               & \acrshort{pzc} transducer     & \acrshort{pzc} transducer     & Inductive transducer                  & $3$E$-3-3\phantom{.00}$E$2$                           & {$E$}, {$Q_\mathrm{E}^{-1}$}                                  & \acrshort{tub}        \\
                                                & \citet{peselnick1979}                         & Granite               & Capacitive transducer         & Capacitive transducer         & Capacitive transducer                 & $2$E$-1$                                              & $E$, $Q_\mathrm{E}$                                           & \acrshort{usgs}       \\
                                                & \citet{riabokon2021a}                         & Sandstone             & \acrshort{pzt} actuator       & Signal generator              & \acrshort{ecp} and \acrshort{lv}      & $5$E$0\phantom{-}-4\phantom{.00}$E$1$                 & $E$                                                           & \acrshort{ua}         \\
                                                & \citet{spencer1981}                           & Various               & Shaker                        & \acrshort{pzt} transducer     & Displacement transducer               & $4$E$0\phantom{-}-4\phantom{.00}$E$2$                 & {$E$}, {$Q_\mathrm{E}^{-1}$}                                  & Chevron               \\
                                                & \citet{sun2018}                               & Sandstone             & Shaker                        & Aluminium STD                 & Strain gauges\tnote{S}                & $1$E$0\phantom{-}-2\phantom{.00}$E$3$                 & {$E$}, {$Q_\mathrm{E}^{-1}$}, {$\nu$}                         & \acrshort{cup}        \\
                                                & \citet{sun2020b}                              & Sandstone             & \acrshort{pzt} actuator       & Aluminium STD                 & Strain gauges\tnote{F}                & $1$E$0\phantom{-}-3\phantom{.00}$E$2$                 & {$E$}, {$Q_\mathrm{E}^{-1}$}, $\nu$, $Q_\upnu^{-1}$           & \acrshort{ens}        \\
                                                & \citet{szewczyk2016}                          & Shales                & \acrshort{pzt} actuator       & \acrshort{pzt} (Kistler)      & Strain gauges\tnote{F}                & $1$E$0\phantom{-}-1.55$E$2$                           & {$E$}, {$\nu$}                                                & \acrshort{ntnu}       \\
                                                & \citet{takei2011}                             & Polycrystals          & \acrshort{pzt} actuator       & Load cells                    & Optical                               & $1$E$-3-1\phantom{.00}$E$1$                           & $E$, $Q_\mathrm{E}^{-1}$                                      & \acrshort{ut}         \\
                                                & \citet{tisato2012}                            & Sandstone             & Motor                         & Load cell                     & Strain gauge cantilever               & $1$E$-1-1\phantom{.00}$E$1$                           & {$E$}, {$Q_\mathrm{E}^{-1}$}                                  & \acrshort{eth}        \\
                                                & \citet{tisato2016}                            & Sandstone             & Motor                         & Load cell                     & Strain gauges                         & $1$E$-1-2.5\phantom{0}$E$1$                           & {$E$}, {$Q_\mathrm{E}^{-1}$}                                  & \acrshort{uoft}       \\
                                                & \citet{tisato2019}                            & Marble                & Motor                         & Load cell                     & Capacitive gap sensors                & $1$E$-1-1\phantom{.00}$E$2$                           & {$E$}, {$Q_\mathrm{E}^{-1}$}                                  & \acrshort{ingv}       \\
                                                & \citet{usher1962}                             & Various               & Vibrators                     & Optical                       & Optical                               & $2$E$0\phantom{-}-4\phantom{.00}$E$1$                 & {$E$}, {$\Delta W/W$}                                         & \acrshort{icl}        \\
                                                & \citet{yee1982}                               & Lucite                & Hydraulic actuator            & Load cell                     & Extensometer                          & $1$E$-2-1.1\phantom{0}$E$1$                           & {$E$}, {$Q_\mathrm{E}^{-1}$}                                  & \acrshort{ge}         \\
                                                & \citet{yurikov2021b}                          & Sandstone             & \acrshort{pzt} actuator       & Aluminium STD                 & Fibre optics                          & $1$E$-1-2\phantom{.00}$E$2$                           & {$E$}, {$Q_\mathrm{E}^{-1}$}, {$\nu$}                         & \acrshort{cu}         \\
    \midrule
    \multirow{9}{*}{\textbf{(ii) Torsional}}    & \citet{behura2007}                            & Unknown               & Spindle                       &                               & Transducer                            & $1$E$-2-8\phantom{.00}$E$1$                           & {$G$}, {$Q_\mathrm{G}^{-1}$}                                  & \acrshort{csm}        \\
                                                & \citet{berckhemer1982}                        & Various               & Electromagnetic drivers       & Inductive transducer          & Inductive transducer                  & $1$E$-3-1\phantom{.00}$E$2$                           & {$G$}, {$Q_\mathrm{G}^{-1}$}                                  & \acrshort{fu}         \\
                                                & \citet{gribb1998a}                            & Olivine               & Motor                         & Transducer                    & Inductive transducer                  & $5$E$-3-1\phantom{.00}$E$1$                           & $G$, $Q_\mathrm{G}^{-1}$                                      & \acrshort{uw}         \\
                                                & \citet{lee2000}                               &                       & Coils                         &                               & Interferometer                        & $1$E$-3-1\phantom{.00}$E$1$                           & $Q_\mathrm{G}^{-1}$                                           & \acrshort{uw}         \\
                                                & \citet{paffenholzetal1989}                    & Various               & \acrshort{pzc} transducer     & Aluminium STD                 & Inductive transducer                  & $3$E$-3-1\phantom{.00}$E$2$                           & {$G$}, {$Q_\mathrm{G}^{-1}$}                                  & \acrshort{tub}        \\
                                                & \citet{peselnick1961}                         & Limestone             & Pendulum                      & Transducer                    & Inductive transducer                  & $4$E$0\phantom{-}-1\phantom{.00}$E$1$                 & {$G$}, {$Q_\mathrm{G}^{-1}$}                                  & \acrshort{usgs}       \\
                                                & \citet{gueguen1989}                           & Forsterite            & Electromagnetic drivers       & Photodiode                    & Photodiode                            & $1$E$-5-1\phantom{.00}$E$1$                           & {$G$}, {$Q_\mathrm{G}^{-1}$}                                  & \acrshort{ipgs}       \\
                                                & \citet{jackson1984}                           & Granite               & Electromagnetic drivers       & Elastic STD                   & Capacitive transducer                 & $1$E$-3-1\phantom{.00}$E$0$                           & {$G$}, {$\delta$}                                             & \acrshort{anu}        \\
                                                & \citet{saltiel2017}                           & Granite               & Electromagnetic drivers       & Load cell                     & Capacitive transducer                 & $1$E$-3-1\phantom{.00}$E$2$                           & {$G$}, {$Q_\mathrm{G}^{-1}$}                                  & \acrshort{lbl}        \\
    \midrule
    \multirow{4}{*}{\textbf{(ii) Flexural}}     & \citet{ikeda2020}                             &                       & \acrshort{pzt} actuators      & Titanium STD                  & Capacitive sensor                     & $1$E${-1}-1\phantom{.00}$E$1$                         & $E_\mathrm{F}$, $Q_\mathrm{E_{F}}^{-1}$                       & \acrshort{uta}        \\
                                                & \citet{jackson2011}                           & Fused silica          & Electromagnetic drivers       & Elastic STD                   & Capacitive transducer                 & $1$E$-3-1\phantom{.00}$E$0$                           & $E_\mathrm{F}$, $\delta$                                      & \acrshort{anu}        \\
                                                & \citet{lee1997}                               & Glass ceramics        & Electromagnetic actuator      & Load cell                     & \acrshort{lvdt}s                      & $1$E$-5-1\phantom{.00}$E$-1$                          & $E_\mathrm{F}$, $\delta$                                      & \acrshort{uw}         \\
                                                & \citet{woirgard1978}                          & Igneous rocks         & Electromagnetic drivers       &                               & Inductive transducer                  & $2$E$-2-1\phantom{.00}$E$1$                           & $Q_\mathrm{E_\mathrm{F}}^{-1}$                                & \acrshort{cnrs}       \\
    \midrule
    \multirow{4}{*}{\textbf{(iv) Volumetric}}   & \citet{adelinet2010}                          & Basalt                & Confining pump                & Pressure sensor               & Strain gauges\tnote{F}                & $1$E$-2-1\phantom{.00}$E$-1$                          & {$K$}, {$Q_\mathrm{K}^{-1}$}                                  & \acrshort{ens}        \\
                                                & \citet{borgomano2020}                         & Limestone             & \acrshort{pzt} actuator       & Aluminium  STD                & Strain gauges\tnote{F}                & $4$E$-3-1.34$E$0$                                     & {$K$}, {$Q_\mathrm{K}^{-1}$}, {$\nu$}                         & \acrshort{ens}        \\
                                                & \citet{mccann2019msc}                         & Water                 & \acrshort{pzt} actuator       & Pressure sensor               & Load cell                             & $1$E$-1-1\phantom{.00}$E$2$                           & $K$, $Q_\mathrm{K}^{-1}$                                      & \acrshort{uta}        \\
                                                & \citet{pimienta2015b}                         & Sandstone             & Confining pump                & Pressure sensor               & Strain gauges\tnote{F}                & $5$E$-3-5\phantom{.00}$E$-1$                          & {$K$}, {$Q_\mathrm{K}^{-1}$}                                  & \acrshort{ens}        \\
    \midrule
    \multirow{2}{*}{\textbf{(v) Uniaxial}}      & \citet{lozovyi2019}                           & Shale                 & \acrshort{pzt} actuator       & \acrshort{pzt} transducer     & Strain gauges\tnote{F}                & $1$E$-1-4\phantom{.00}$E$0$                           & {$C_{33}$}                                                    & \acrshort{ntnu}       \\
                                                & \citet{suarez2001}                            & Shale                 &                               &                               &                                       & $1$E$0\phantom{-}-2\phantom{.00}$E$2$                 & {$C_{33}$}                                                    & SINTEF                \\
    \bottomrule[1.5pt]
    \end{tabular}}%
    \begin{tablenotes}\tiny
        \item[S] is for \textit{S}emiconductor strain gauges.
        \item[F] is for \textit{F}oil strain gauges.
    \end{tablenotes}

\end{threeparttable}
\end{sidewaystable}

\subsection{OTHERS}

\begin{figure}[H]
    \centering
    \includegraphics[width=0.7\textwidth]{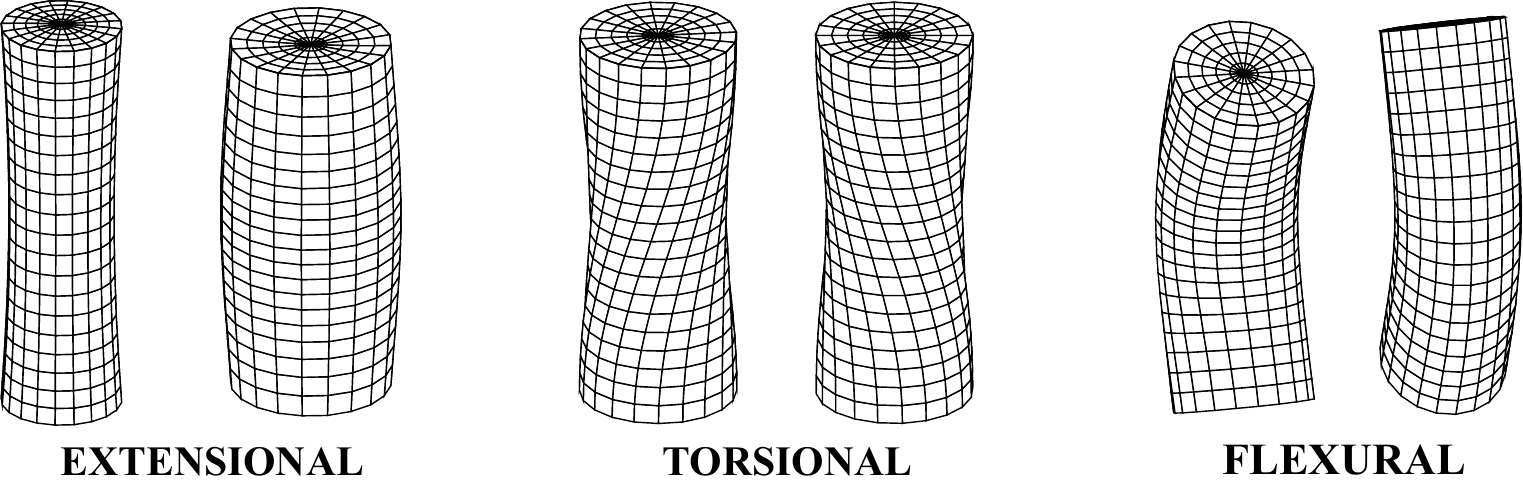}
    \caption{Exaggerated surface particle displacement for extensional, torsional, and flexural modes for \acrshort{rus} (but also applies to \acrshort{fo}) measurements. Modified from \citet{zadler2004}.}
    \label{fig:zadler}
\end{figure}

Beyond the traditional \acrshort{rb}, \acrshort{pt}, and \acrshort{fo} techniques, \acrfull{rus}, \acrfull{dars}, and \acrfull{lus} are other novel techniques exploited in rock physical research. \citet{maynard1996} not only introduced the term \acrshort{rus} \say{to encompass all techniques in which ultrasonic resonance frequencies are used to determine elastic moduli} but also traced the history of \acrshort{rus} back to \citet{fraser1964,schreiber1970,demarest1971}. In fact, \acrshort{rus} is in many aspects a continuation of \acrshort{rb} but differs due to its ability to capture numerous resonance peaks across wider frequencies irrespective of material shape and dimensions. Be it extensional, torsional, and flexural modes, \citet{ulrich2002,zadler2004} describe \acrshort{rus} as a technique able to ascertain the suite of elastic moduli for anisotropic rocks during a single frequency sweep (Figure~\ref{fig:zadler}). For orthorhombic symmetry, eight independent groups of free-vibration modes containing all possible combinations of these symmetries (including isotropic, cubic, hexagonal, and tetragonal) exist: one extensional and torsional, plus three shear and flexural \citep{demarest1971,ohno1976}. \acrshort{rus} is as numerical as it is experimental in the sense that (i) forward and (ii) inverse problems need be solved to validate the experiment: (i) compute the frequencies and shapes of the normal modes, and (ii) apply a non-linear inversion algorithm to find the elastic constants from these normal-mode frequencies. Since \acrshort{rus} detects numerous resonance peaks from all eight groups, exact correspondence between resonance peaks and vibration modes is paramount as any dissonance makes for erroneous measurements. \citet{batzle2006} remarked on \acrshort{rb} versus \acrshort{rus} that the narrow bandwidth of \acrshort{rb} limits the probed frequencies to the primary resonance and a few overtones whereas the extended frequencies of \acrshort{rus} renders it susceptible to jacketing and suspension procedures plus inhomogeneous strain conditions for the torsional and flexural modes. Indeed, \citet{zadler2004} acknowledged that \textit{in-situ} measurements involving strong coupling to a pressurized medium violate the stress-free boundary conditions \acrshort{rus} is based upon. 

\citet{harris1996} envisioned \acrshort{dars} to measure the change in resonance frequency of a submerged cavity caused by the absence or presence of a foreign object inside the cavity which perturbs the resonance properties of the cavity. \acrshort{dars} is based on perturbation theory which relates the frequency shift between a cavity with or without a specimen to the acoustic properties of said specimen \citep{harris2005,wang2012,vogelaar2015}. In other words, \acrshort{dars} is restricted to the determination of bulk modulus as the change between the normal modes of two volumetric effects. Keys to this technique are (i) the cavity immersed in a fluid containing vessel, (ii) piezoceramic sources used to excite fluid resonances, and (iii) a hypersensitive hydrophone embedded on the cavity surface for detection of acoustic pressure signal changes. Added to these three key elements are also (iv) a computer-controlled step motor for accurate and repeatable specimen positioning, and (v) a phase-sensitive lock-in amplifier connected to source and receiver that uses a predefined frequency sweep (around the natural resonance of the cavity) to select the resonance curve of the fundamental mode in order to recognize the received signal at a specific reference frequency and phase. \citet{wang2012} implemented a \acrshort{fem}-based simulation to better understand the \acrshort{dars} system and improve its accuracy in estimating acoustic properties. 

\citet{scruby1980} first described \acrshort{lus} as a technique to \say{study thermoelastic generation of elastic waves in a metal by unfocused laser radiation.} As such, \acrshort{lus} is able to characterize waves in terms of velocity and attenuation. Instead of relying on mechanical coupling between transducer and specimen like \acrshort{pt} and \acrshort{fo}, \acrshort{lus} is strictly non-contact due to two lasers generating and recording elastic waves at the specimen surface from afar. A short pulse of electromagnetic radiation delivered by the first laser causes thermoelastic-expansion-induced ultrasonic waves recorded by the second laser (interferometer) at an arbitrary point on the specimen surface \citep{scruby1990}. Since the surface area beamed by the laser is significantly smaller than the area coupled to \acrshort{pzt} transducers, \acrshort{lus} enables multidirectional characterisation of waves from a single specimen while also guaranteeing that group velocity is unambiguously measured. The eternal group versus phase velocity conundrum is thus avoided as the former is easily converted to the latter using established algorithms. It is also claimed that the non-contact generation and detection is not frequency-limited by the physical dimensions of the transducer elements \citep{pouet1990,pouet1993,simpson2019} yet it is curious that this possibility remains unexplored as most studies are limited to ultrasonic frequencies (thus the name) considering that seismic frequencies are alpha and omega in practice. If \acrshort{lus} is indeed universal for all frequencies, \acrshort{rb}, \acrshort{pt}, and \acrshort{fo} would all be redundant. However, even if \acrshort{lus} is theoretically able but practically unable to probe all frequencies, it is still an improvement compared to \acrshort{pt} because \acrshort{lus} leaves less ultrasonic frequencies unprobed. In the realm of rocks, \acrshort{pt} measured parameters at ultrasonic frequencies are often regarded to coincide with the high-frequency limit but this need not be true and could be further explored by \acrshort{lus}. For example, there is evidence for ultrasonic attenuation in rocks depending on the enforced conditions (e.g. \citet{johnston1980}). \acrshort{lus} is extended from metals to rocks in the absence and (recently) in the presence of \textit{pseudo in-situ} conditions \citep{scales2003,lebedev2011,blum2013,carson2014,adam2014,xie2016,adam2017,xie2018,simpson2019,simpson2020,adam2020,simpson2021,simpson2022}. \textit{Pseudo} is emphasized since it is still restricted to isotropic stress conditions but remain unrestricted in terms of temperature \citep{simpson2019,simpson2020,simpson2021}.


\section{DISCUSSION}

Technique-bias currently exists because several different techniques need be exploited to access Hz, kHz, and MHz frequencies. These techniques also measure different parameters that need be transformed for comparability which is in itself is an error source. Only if theoretically understood in a controlled environment (laboratory) may dispersion also be practically understood in an uncontrolled environment (field). To the end of understanding dispersion as a phenomenon, all frequencies need be accessible for the transitions between regimes to be studied. These transitions are mechanism-dependent and observed as changes in moduli or attenuation peaks with frequency. Dispersion and attenuation are indeed causality-bound. They are also key to identify and understand the physical mechanisms causing dispersion. Multiple transitions probably occur in any given rock but singular transitions are mostly studied due to frequency-limitations of the investigation techniques. The limiting factors for \acrshort{fo}, \acrshort{rb}, and \acrshort{pt} are resonances (upper boundary), specimen size, and piezoceramic size, respectively. Resonances are constructive for \acrshort{rb} and destructive for \acrshort{fo} but the same principle applies to both: The smaller (and lighter) the resonator, the higher the characteristic frequency, and vice versa. In fact, this principle also applies to \acrshort{pt} where larger piezoceramics equal lower frequency. \acrshort{pt} piezoceramics for this purpose seldom recede below $\sim50$ kHz. No universal technique probing all frequencies of interest alas exists at this time. As a result, existing techniques are enhanced or manipulated to extend their operating frequencies. For example, \acrshort{fo} apparatuses are becoming smaller and lighter to increase the upper boundary at which they resonates. Numerical investigations during the design process accelerated this evolution. \citet{batzle2006} identified nodes and antinodes to extend this upper boundary by distributing the measurements accordingly. Since dispersion is a fluid-dependent phenomenon, viscosity-manipulation either by fluids such as glycerine \citep{batzle2006,david2012,subramaniyan2015,pimienta2015a,pimienta2015b,yin2017,borgomano2019,gallagher2020,han2021,ma2021} or temperature \citep{jones1983,batzle2006,spencer2013,rorheim2021a} may shift the transition from unprobed to probed frequencies due to altered fluid mobility. Viscosity-manipulation by temperature \citep{batzlewang1992} is explored at high but unexplored at low temperatures for \acrshort{fo}. Among \acrshort{rus}, \acrshort{lus}, and \acrshort{dars}, particularly \acrshort{lus} is enticing due to its non-contact nature and its supposed ability to \say{generate acoutic energy} and \say{detect surface displacements over a broad frequency band (typically from continuous to GHz)} \citep{pouet1990}. Even if \acrshort{lus} is unable to probe seismic frequencies it is proven to probe a wider range of ultrasonic frequencies than \acrshort{pt}. Thus far only isotropic (hydrostatic) conditions are possible \citep{simpson2019} but applying a load along the axis of two synchronized stepper motors above and below the specimen to achieve biaxial conditions while still preserving its rotation ability should be fairly simple. Since most of the sensors are outside the pressure vessel, it is also an ideal candidate to fit inside a \acrshort{ct} scanner provided that the laser signals are not somehow distorted by x-rays. \say{A respectable frequency gap remains to be bridged} \citep{birch1938} are still true words that are becoming less so with time and technological advancements.

Models compensate for the unprobed frequencies and also connect different measurements at different frequencies. Commonest amongst a multitude of models are \acrshort{cc}, \acrshort{sls}, and \acrshort{kkr}. \acrshort{kkr} is superior to \acrshort{cc} and \acrshort{sls} due to its analytical nature but requires parameters (real part) and their corresponding attenuations (imaginary part) at all frequencies to be valid. Many \acrshort{wiff} models based on \citet{biot1956a} exist because \say{squirt-flow} is the dominant dispersion mechanisms at microscopic scale in sandstones. Models simulating dispersive behaviour related to mechanisms valid at mesoscopic and macroscopic scales are rarities in comparison. Since multiple transitions may occur in the same specimen, it is important to simulate mesoscopic and macroscopic flows in addition to microscopic flow. As described in Section~\ref{sec:methods}, not only analytical but also numerical models are developed to describe these mechanisms. The paucity of bound water models is concerning because it is perhaps the least known yet also the most critical mechanism due to its experimentally \citep{antognozzi2001,major2006,goertz2007,ulcinas2011} and numerically \citep{holt2017} proven non-zero shear modulus and enhanced viscosity. Simultaneous bound water conversion to free water and layer thickness reduction are also proven at elevated temperatures \citep{zymnis2018}. Bound water is also proven to convert to free water and  \acrshort{drp} is progressing and may be a valuable tool in the future. Calibrating \acrshort{drp} with $3$D printed or otherwise synthetic specimens with known geometry and properties would surely be a great leap forward if it becomes a possibility. More knowns than unknowns is key.

\citet{subramaniyan2014} is also a precautionary tale that discusses potential problems and possible solutions with implementing \acrshort{fo}. As such, advantages and disadvantages with different components as well as boundary effects and strain amplitudes are discussed. \citet{rorheim2019} discovered that modulus is unaffected but attenuation is affected by minor misalignments as foreseen by \citet{liu1983}: smooth and parallel surfaces are paramount. For longitudinal excitation, deviatoric stress $\sigma_\mathrm{dev}=\sigma_\mathrm{ax}-P_\mathrm{c}$ (where $P_\mathrm{c}$ is the previously undefined confining pressure) can be a mitigation measure to ensure adequate coupling and uniform stress distribution. Other mitigation measures include adhesive \citep{spencer1981} and aluminium foil \citep{li2017} at the specimen-endcap interfaces. \acrshort{fo} apparatuses differ in primarily strain and secondary stress measurements (Table~\ref{tab:FO}). Most measure local strains using a set of specimen-attached strain gauges while few measure bulk strain using capacitors, cantilevers, \acrshort{lvdt}s, lasers, or recently even fibre optics \acrshort{das} sensors. \textit{Pro et contra} of bulk versus local measurements are elaborated (Section~\ref{sec:methods}) where bulk is considered superior to local (especially for heterogeneous rocks) but more problematic for radial measurements due to the double-interface problem. Strain gauges are more directional-versatile and require less calibration than the others but must be mounted to each individual specimen where the wires cause sealing-related problems. \acrshort{das} is particularly interesting as it offers the opportunity to bridge the laboratory-field void because equivalent sensors are used in both environments. Mechanisms can thus be studied independent of potential sensor-bias. \acrshort{das} is also more strain-sensitive (as low as $10^{-11}$ strain amplitudes) and less temperature-sensitive than semiconductors among several advantages \citep{yurikov2021b}. Elastic standards are customary among numerous stress sensors. These standards are commonly non-dispersive metals such as aluminium or titanium attached with foil or semiconductor strain gauges. Due to the high stiffnesses of both metals, high-sensitive semiconductors are preferred over low-sensitive foil strain gauges. Stress $\sigma_\mathrm{ax}$ and strain $\epsilon_\mathrm{ax}$ proportionally increase while \citeauthor{young1807}'s modulus $E$ is constant (Equation~\ref{eq:young}) if the area $A$ is lowered since $\sigma_\mathrm{ax}=F/A$ where $F$ is force. Increasing the strain amplitudes of the standard improves the accuracy of foil strain gauges in regard to amplitude and phase sensitivities. For example, \citet{rorheim2022phd} designed a temperature-compensated \say{hollow dog bone} aluminium standard equipped with four biaxial foil gauges where $A$ was decreased by a factor of five. Note that electronic-induced errors are a concern if different circuitries are used for stress and strain measurements \citep{batzle2006,madonna2013,szewczyk2016}.

\nomenclature{$\sigma_\mathrm{dev}$}{Deviatoric stress.}
\nomenclature{$P_\mathrm{c}$}{Confining pressure.}
\nomenclature{$A$}{Area.}
\nomenclature{$F$}{Force.}

\acrshort{fo} as a measure to probe seismic frequencies is a force to be reckoned with manifested by the growing number of apparatuses. Most of which are based on the design of \citet{batzle2006} with specimen-attached strain gauges for local strain measurements opposed to specimen-unattached transducers for bulk strain measurements like \citet{spencer1981}. Simultaneous local and bulk strain measurements combined in one apparatus could possibly distinguish between microscopic, mesoscopic, and macroscopic mechanisms as dispersion and attenuation sources \citep{chapman2018}. Other possible \acrshort{fo} improvements for the future include:

\begin{itemize}
    \item Numerical-based designs using, for example, \acrfull{comsol} or \acrfull{moose} framework are crucial in understanding the nature of resonance and predicting their nodal and antinodal distributions with frequency. Be it \acrshort{drp} or otherwise, numerical studies complementing experimental ones are also advancing the universal understanding of dispersion and attenuation mechanisms.
    \item \citet{tisato2016} was the first of its kind but others will surely follow because the \acrshort{ct}-\acrshort{fo} combination may elucidate the spatial distribution of fluids and thus also the fluid-solid interactions that are the dispersion and attenuation causing mechanisms. Most phenomena and their respective mechanisms are explained if porosity is adequately described as it is the critical factor separating solid and porous media. Measuring physical properties while simultaneous imaging phase distributions at pore-scale is common for other techniques \citep{noiriel2022} except \acrshort{fo}. $3$D printed or otherwise synthetic rock specimens with known geometry and properties subject to \acrshort{ct} plus \acrshort{fo} could also contribute to this end. Fractals may also be experimentally investigated if specimen geometries are scale-independent and fractal-consistent.
    \begin{itemize}
        \item Imaging fluid-solid interactions with time is especially enticing for \acrfull{ccs} purposes (as a natural extension to \citet{rorheim2021c}).
    \end{itemize}
    \item Multi-element \acrshort{pzt} transducers could also excite torsional and flexural modes as well as mitigate any potential misalignments by redistributing the stress field. \citet{nakagawa2013LF} partially converted longitudinal into torsional excitations using a novel \say{compression-torsion coupler}. Misalignment mitigation and stress field redistribution is already achieved by employing three \acrshort{pzt} transducers \citep{paffenholzetal1989,takei2011,ikeda2020}.  
    \item True triaxial conditions ($\sigma_\mathrm{x}\neq\sigma_\mathrm{y}\neq\sigma_\mathrm{z}$) \citep{walle2019a,walle2019b} are achieved if there is independent control of two radial strains in which $H=C_{11}$ in Equation~\ref{eq:direct_p}. Pistons in addition to confining pressure are required to achieve these conditions which render specimen-attached strain gauges inadequate due to the fact that they would be destroyed upon impact of the piston. A cantilever solution in both axial and radial directions could however be a possibility.
    \item Virtual instead of physical lock-in amplifiers \citep{yao2013,yao2013phd} would offer greater data control as well as being an inexpensive option to split strain gauge elements to focus on local pore-scale processes by analyzing individual strains before averaging to distinguish local from global flow.
    \item Temperature control by heating circulators would render experimental temperature independent from laboratory temperature which is sensitive. Improved temperature control would not only allow for unexplored (sub- as well as superambient) temperatures to be explored but also keep the temperature constant for the duration of the experiment. Subambient temperatures are an unexplored curiosity shifting the mechanism-dependent transitions towards lower frequencies due to decreased fluid mobility.
    \item Experiments dedicated to elucidate bound water as an attenuation mechanism are crucial to understand attenuation as a phenomenon. \acrshort{fo} is exploited to this end as well albeit still in its infancy \citep{rorheim2021a,long2022}. Other novel techniques or combinations of techniques may however be more applicable.
    \item \acrfull{ml} is bound to somehow be exploited in the laboratory as it is in the field \citep{zhu2018} to ease data acquisition as well as analysis. \acrshort{pt} is perhaps favoured over \acrshort{fo} though as automated arrival time picking \acrshort{ml} is an immediate extension of \citeauthor{zhu2018}'s \texttt{PhaseNet}. Especially S-waves could benefit from automated picking due to their ambiguity compared to P-waves. \acrshort{ml} could also possibly be combined with \acrshort{drp}.
    \item Exploring higher harmonics is not only possible for \acrshort{rb} at sonic but also for \acrshort{fo} at seismic frequencies. Harmonics are key to understand non-linearities and non-elastic mechanics and their dependency on strain amplitude, frequency, pressure, and fluid saturation \citep{tutuncu1998,riviere2016,mews2022}. Nonlinear attenuation is also temperature-dependent: mesoscopic $f_\mathrm{s}$ increases and macroscopic $f_\mathrm{b}$ decreases with temperature \citep{deng2022}. The static-dynamic property discrepancy is caused by mechanisms such as drainage conditions, dispersion, heterogeneities, and strain amplitude \citep{fjaer2019}. Strain amplitude dependency on non-linearities and non-elastic mechanisms are crucial to resolve this discrepancy.
    \begin{itemize}
        \item Pressure- and temperature-induced non-linear elastic behaviour in rocks is also explored by \acrshort{lus} \citep{simpson2021,simpson2022}.
    \end{itemize}
\end{itemize}

\nomenclature{$\sigma_\mathrm{x}$}{One of the principal stresses in $z$-direction.}
\nomenclature{$\sigma_\mathrm{y}$}{One of the principal stresses in $y$-direction.}
\nomenclature{$\sigma_\mathrm{z}$}{One of the principal stresses in $z$-direction.}


\section{CONCLUSION}

Novel experimental techniques need be developed or enhanced to improve the conditions at which rocks are experimented for the field-laboratory gap to be bridged. Among the usual techniques, \acrshort{fo} is perhaps the current champion due to its similarities with field seismic. Despite its promise, caution need be exercised when comparing laboratory with field data due to the omnipresent upscaling problem that transcends frequencies. Novel are not only the experimental techniques but also the analytical and numerical models that are continuously being developed. These models primarily describe \acrshort{wiff} at microscopic scale but secondary models to describe flows at mesoscopic and macroscopic are required for full understanding since multiple transitions are measured in rocks. Bound water should also be further explored as a mechanism due to its experimentally and numerically proven non-zero shear modulus and enhanced viscosity. \acrshort{drp} is gaining momentum and may possibly be integral in understanding rock phenomena and mechanisms in the future. 


\section*{ACKNOWLEDGEMENTS}

Rune M. Holt and Andrew J. Carter at \acrshort{ntnu} are acknowledged for their encouragements. Current and former SINTEF employees are also acknowledged: Andreas Bauer, J{\o}rn F. Stenebr{\aa}ten, Serhii Lozovyi, Dawid Szewczyk, Lars Erik Walle, Andreas N. Berntsen, Eyvind F. S{\o}nsteb{\o}, M. Hossain Bhuiyan, Anna M. Stroisz, Pierre R. Cerasi, and Sigurd Bakheim. Lara Blazevic is finally acknowledged for proofreading. 

\printbibliography[title={BIBLIOGRAPHY}]


\setlist[description]{leftmargin=!, labelwidth=5em}
\printglossary[type=\acronymtype,title=ACRONYMS]
\setlist[description]{style=standard}


\setlength{\nomlabelwidth}{5em}
\renewcommand{\nomname}{NOMENCLATURE}
\newcommand{\nomunit}[1]{%
\renewcommand{\nomentryend}{\hspace*{\fill}#1}}
\printnomenclature

\end{document}